\documentclass[12pt]{article}
\usepackage{fullpage}
\usepackage{amsmath,amssymb}
\usepackage[dvips]{graphicx}
\usepackage{subfigure}
\usepackage{latexsym}
\usepackage{xcolor,graphicx}
\renewcommand{\d}{\text{d}}
\newcommand{\lsuper}[1]{ {}^{#1}\hspace{-2 pt}}

\newcommand{\half}{{\frac{1}{2}}}
\newcommand{\beq}{\begin{equation}}
\newcommand{\enq}{\end{equation}}
\newcommand{\bs}{\boldsymbol}

\begin{document}

\begin{center}
{\large \bf Quantum mechanics as a statistical description of\\ classical electrodynamics}\\

\medskip
Yehonatan Knoll \\
%

\begin{abstract}
\noindent It is shown that quantum mechanics is a plausible       statistical description of an ontology described by classical electrodynamics. The reason that no contradiction arises with  various no-go theorems regarding the compatibility of QM with a classical ontology, can be traced to  the fact that classical electrodynamics of interacting particles has never been given a consistent definition.  Once this is done, our conjecture follows rather naturally, including a purely classical explanation of photon related phenomena. Our analysis entirely rests on the block-universe view entailed by relativity theory.  
\end{abstract}

\end{center}

\section{Introduction}
We have all been taught that classical electrodynamics (CE) of point charges cannot be an  ontology underlying the statistical predictions of quantum mechanics (QM). Classical particles, we were told, cannot diffract or tunnel, nor can classical atoms exist, as their electrons would quickly dissipate their energy into radiation and spiral towards the nucleus. We were further educated that CE can't explain Compton's and the photoelectric effects, and that Bell's inequality is incompatible with a CE ontology. But what our teachers forgot to mention is that CE of point charges is ill defined due to the so-called classical self-force problem---the singularity of the electromagnetic (EM) field on the world line of a point charge, rendering the Lorentz force ill defined, and the  EM energy non-integrable. How, then, can an ill defined theory be incompatible with anything, QM in particular?

There are, of course, many versions of CE which are well defined. Ignoring the (ill defined) self-force leads to a well defined CE; Extending the point charge distribution to a small sphere leads to another. But each of these methods results in a distinct theory. Which of them \emph{is} CE---that classical theory which, so we were told, is so manifestly incompatible with quantum phenomena?
 
To answer this question we first need ask a related question: How is it that an ill defined theory can be such a practical---often very accurate---tool, as CE evidently is? The answer given to this question in \cite{YK2016} is that the results of CE, in all those cases where it is successfully applied, can be extracted from a  small set of tenets, referred to as the \emph{basic tenets of CE}, which are not only sufficient but also necessary to explain \emph{all} those experimental results. 

It turns out, however, that all previous attempts to render CE well defined  fail to completely satisfy the basic tenets and not because their  importance was overlooked.  Dirac's \cite{Dirac} celebrated solution to the classical self-force problem, for example,  is explicitly motivated by them but fails to deliver. In a recent paper  \cite{YK2016}  by the current author the task of rendering CE well defined subject to the basic tenets  has been accomplished, providing a proof for their self-consistency. That is, no additional `glue', which might have rendered the basic tenets invalid at the microscopic level, is needed in order to describe finite size charged particles.

There are, however, many cases in which a well defined CE respecting the basic tenets is not enough to model an experiment; cases in which it is not specific solutions of CE which are of interest, but rather statistical aspects of ensembles of such solutions. The composition of the ensemble, in most cases, cannot be derived from  CE  plus some obvious assumptions, such as a uniform distribution over the impact parameter in scattering experiments. Nor are there obvious  ways to even express the ensemble, as in the case of statistical mechanics (irrespective of the fact that the ergodicity postulate cannot be defended in all cases). The composition of the ensemble must be regarded as a \emph{complimentary, fundamental law of nature}, on equal footing with CE.  In the current paper we analyse a generic such complementary  statistical law, assumed to describe ensembles of CE solutions. QM, we argue, emerges naturally this way.

\section{The basic tenets of CE}\label{ECD}

The two obvious pillars of CE  are Maxwell's equations in the presence of a conserved\footnote{Solutions to Maxwell's equations exist for a conserved source only.}  source due to all particles (labelled by $a=1\ldots n$)
\begin {equation}\label{Maxwell's_equation}
  \partial_\nu F^{\nu \mu} \equiv \partial^\nu\partial_\nu A^\mu - \partial^\mu\partial_\nu  A^\nu = \sum_a\,j^{(a)\,\mu} \, ,
\end {equation}
\beq\label{continuity_equation}
\partial_\mu j^{(a)\,\mu}=0\,,
\enq
with $F_{\mu \nu}=\partial_\mu A_\nu - \partial_\nu A_\mu$ the antisymmetric Faraday tensor, and  local `Lorentz force equation'
 \beq\label{ni}
\partial_\nu\,T^{(a)\,\nu \mu} =F^{\mu\nu}\,j^{(a)}{}_\nu\,,
\enq
with $T^{(a)}$ the `matter' e-m tensor associated with particle $a$.
Defining the \emph{canonical tensor}   
\begin {equation}\label{Theta}
\Theta^{\nu \mu}=\frac{1}{4} g^{\nu \mu} F^{\rho\lambda}F_{\rho\lambda} + F^{\nu \rho}F_\rho^{\phantom{1} \mu}\,,
\end {equation}
we get from \eqref{Maxwell's_equation} and \eqref{Theta}  Poynting's theorem
\begin {equation}\label{Poynting}
 \partial_\nu \Theta^{\nu
 \mu} =- F^{\mu}_{\phantom{1}\nu}\sum_a\,j^{(a)\,\nu}\,.
\end {equation}
Summing \eqref{ni} over $a$ and adding to \eqref{Poynting} we get local e-m conservation
\beq\label{pp}
 P=\Theta + \sum_a\,T^{(a)}\quad\Rightarrow \quad \partial_\nu P^{\nu\mu}=0\,, 
\enq
and, purely by the symmetry and conservation of $P^{\nu\mu}$, also generalized angular momentum conservation
\beq\label{angular_momentum}
\partial_\mu {\mathcal J}^{\mu\nu\rho}=0\,,\quad {\mathcal J}^{\mu\nu\rho}=\epsilon^{\nu\rho\lambda\sigma}P^\mu_{\phantom{1}\sigma}x_\lambda\,.
\enq

As shown in \cite{YK2016}, if $j^{(a)}$ and $T^{(a)}$ associated with each particle are co-supported on a  world-line, corresponding to `point-particle' CE, no realization of the basic tenets can be found. A valid realization must therefore involve $j$ and $T$ extending beyond this support yet still localized about it, representing what can be called `extended particles' with  non-rigid internal structures. Nevertheless, the reader must not take too literally this name, as both $j$ and $T$ associated with distinct particles are allowed to overlap or cross one another.

{\bf Symmetries.} Implicit in the basic tenets \eqref{Maxwell's_equation}--\eqref{Theta} is their  Lorentz covariance. This symmetry group can be extended to include dilatations of space-time by some constant $\lambda>0$, by postulating the following transformation law, preserving the basic tenets: 
\beq\label{scale covariance}
A(x)\mapsto\lambda^{-1}A\left(\lambda^{-1}x\right)\,,\quad j(x)\mapsto\lambda^{-3}j\left(\lambda^{-1}x\right)\,,\quad T(x)\mapsto\lambda^{-4}T\left(\lambda^{-1}x\right)\,,  
\enq
Scale covariance or, equivalently, the absence of a privileged scale, is just as natural a requirement as translation covariance, if a theory has any chance of becoming a fundamental description of reality. The mass (scale) of a particle appears therefore as an attribute of a specific solution rather than a parameter of the theory, and in section \ref{The ground state} we briefly speculate about the origin of such `spontaneous symmetry breaking' giving all particles of the same specie their common mass.

Whereas for Lorentz boosts and rotations, the Lorentz transformed  tensors are readily associated with a transformed physical reality by means of a corresponding (passive) Lorentz transformation of the reference frame, it is not a priori clear what physical reality corresponds to the remaining two Lorentz transformations: `Time reversal' $\text{T}:=\text{diag}(-1,1,1,1)$ and `space reflection' $\text{P}:=\text{diag}(1,-1,-1,-1)$, nor is it even clear that any physical reality \emph{should} correspond to them.  It turns out, for example, that only the product PT (or `CPT', as the sign of the charge is changed )   
\beq\label{PT}
A(x)\mapsto -A(-x)\,,\quad j(x)\mapsto -j(-x)\,,\quad T(x)\mapsto T(-x)\,,
\enq
is respected by the realization in \cite{YK2016} implying, among else, a very specific sort of antiparticle.
     
{\bf Arrow-of-time.}
In a universe in which no particles imply no EM field, a solution of   Maxwell's equations is  uniquely determined by the conserved current, $j$. The most general such dependence which is both Lorentz and gauge covariant takes the form  
\begin {equation}\label{convolution with K}
 A^\mu(x)=\int d^4 x' \big[\alpha_\text{ret}(x') K_\text{ret}{}^{\mu \nu}(x-x') + \alpha_\text{adv}(x')  K_\text{adv}{}^{\mu \nu}(x-x')\big]j_\nu(x')\,, 
\end {equation}
for some (Lorentz invariant) space-time dependent functionals, $\alpha$'s, of the current $j$, constrained by  $\alpha_\text{ret}+\alpha_\text{adv}\equiv1$, where  $K_{{}^\text{ret}_\text{adv}}$ are the advanced and retarded   Green's function of \eqref{Maxwell's_equation}, defined by \footnote{More accurately, \eqref{K defined} and \eqref{causality condition} do not uniquely define $K$ but the remaining freedom can be shown to translate via \eqref{convolution with K} to  a gauge transformation $A\mapsto A +\partial \Lambda$, consistent with the  gauge covariance of  ECD.}
\begin {equation}\label{K defined}
\left(g_{\mu \nu} \partial^2 - \partial _\mu \partial_\nu\right) K_{{}^\text{ret}_\text{adv}}{}^{\nu
\lambda}(x)=g_\mu^{\phantom{1}\lambda} \delta^{(4)}(x)\,,
\end {equation}
\begin {equation}\label{causality condition}
 K_{{}^\text{ret}_\text{adv}}(x)=0\,\,\,\ \text{for }\, x^0 \lessgtr 0\,.
\end {equation}

The standard $\alpha_\text{adv}\equiv0$ proviso, not only is it likely to be incompatible with an otherwise valid realization of the basic tenets (such as the one in \cite{YK2016}) but, moreover, it is not implied by observations. The observed asymmetry between retarded and advanced fields---the so-called radiation arrow-of-time---only applies to macroscopic systems undergoing an irreversible process, such as a burning candle. Moreover, in such processes it is only the \emph{difference} between the integrated retarded and advanced Poynting fluxes across a surface surrounding the system, to which the rate of e-m conversion equals. Put differently, the entire retarded Poynting flux is never directly measured, even in seemingly classical contexts, such as in the detection of radio waves. As we shall see, radiation phenomena associated with any single particle, involve the rest of the particles in the universe as well. `Running the movie backwards', thereby reversing the arrow-of-time, without changing \emph{all} particles to their antiparticles, is therefore forbidden, and the preferred time direction could very well be linked with the particle-antiparticle imbalance in the universe.

As we shall see, radiation phenomena associated with any single particle, involve the rest of the particles in the universe as well. `Running the movie backwards', thereby reversing the arrow-of-time, without changing \emph{all} particles to their antiparticles, is therefore forbidden, and the preferred time direction could very  well be linked with the particle-antiparticle  imbalance in the universe.\\ 
\linebreak
\noindent A concrete realization of the above tenets, first appearing in \cite{KY} and then fine-tuned and related to the self-force problem  in \cite{YK2016}, is dubbed extended charge dynamics (ECD), but in the rest of the paper, unless otherwise stated, we shall use this name to designate a generic realization of the basic tenets by means of extended particles.

\section{The block universe}\label{The block universe}
In its greatest generality, ECD provides a rule for filling empty space-time with energy and momentum. A typical such e-m distribution is concentrated around world lines associated with particles, and in the vicinity of light cones with apexes on those  world lines, corresponding to radiative processes. This rule permits a very restricted, yet infinite  set of  such e-m distributions, one of which  allegedly describes our universe. It is crucial to note that, while  some features are common to all e-m distributions permitted by ECD, others are unique to the specific one filling our universe and, therefore, ECD alone is an incomplete description of the universe.

This view of the universe, as a four dimensional `block' filled with some global $4D$ e-m distribution, goes by the name ``the block universe''. It is the \emph{only} way to make sense of SR, let alone of GR, but this fact seems to be camouflaged by the same mathematical accident which has led generations of physicists to reject advanced solutions of Maxwell's equations (see section 3.3 in \cite{YK2016}). Historically, most (relativistic) block-universe models  were  constrained by some local, Lorenz covariant second order differential relation. The hyperbolicity of Lorentz covariant operators, such as the d'Alembertian,  enables one to specify data on a single space-like surface of dimension three, and use the remaining coordinate to `propagate' the solution to neighboring three-surfaces. Again,  resorting to the erroneous analogy with physical wave equations, that extra propagation parameter is given the meaning of `time'. This offers an (illusion of-) escape route from  fatalism, entailed by the block-universe view: The future is not yet determined; I can still influence it by, say, incorporating into the evolution equations a time-varying source or a force-term. 

Identifying the above mathematical evolution parameter with absolute Newtonian time is clearly meaningless, as it is not unique. Minimizing its role to some personal time, is still wrong. It is just some---again, not unique---coordinate, used in the construction of a four-dimensional object. Eventually, `I' am represented by some four-dimensional distribution of e-m, localized around some world line, with no hint whatsoever as to why `I' feel that `time flows in one direction', or why it is that I seem to be experiencing the block-universe `one space-like slice at a time'. 

Nowadays there are quite a few quantum foundationalists who try to tackle the questions of free-will and the flow-of-time, but it is the firm belief of the current author that such subjective phenomena lie outside the (extremely narrow) realm of physics. A similar tension between determinism and free-will exists in Newtonian physics, and it never prevented physicists from  doing the `simple' things they were meant to do: Build machines and predict the future. In the next section we argue that nothing in the block-universe philosophy should prevent quantum physicists from continuing that tradition. 
\subsection{Classical mechanics emerging in the block-universe}
A general block-universe needs not admit a formulation in terms of a Cauchy initial value problem (IVP), as, historically was the case. That is, no knowledge of data on a space-like surface, including `hidden', unmeasurable data, is sufficient to uniquely determine the data on neighboring surfaces. ECD's realization of the constittive relations in \cite{YK2016} (currently the only known well defined one) is such a non-IVP theory. Nevertheless, in many situations, the existence of  \emph{local} basic tenets, constraining the ECD block-universe, despite a global, non-IVP construction,  leads to a description of the interesting attributes of a system by means of (effective) local differential equations. For example, when shooting a hockey pack---what else can the pack do but keep moving straight if it is to conserve momentum? More generally, it is shown in appendix D of \cite{YK2016} that  when a charged body is moving in a weak  external EM field which is  slowly  varying over the extent of the body, its path is described by solutions of the Lorentz force equation in that field.   \emph{It is these local constraints, despite a global construction, to which classical mechanics  owes its phenomenal power}, enabling the design of predictable, local `machines', whose proper function is indifferent to the  rest of the block universe, but it must be stressed, that such machines are the exception rather than the rule. For example, when an extended particle is moving in an external field which is either strong, or else rapidly varying over its extent,  the construction in appendix D of \cite{YK2016} breaks down and no differential equation can describe the path of the particle anymore. This is presumably the situation of a particle passing through the strong magnetic gradients in a Stern-Gerlach polariser, hence the apparent `indeterminism': Even when we `feed' the polariser with seemingly identically polarized particles (arriving from a previous polariser) we typically get both `spin up' and `spin down' coming out of it.\footnote{If one accepts that  particle-in-a-polarimeter type processes are part of the microscopic description of any complex system, then there is no escape from the conclusion that any macroscopic system classified as `chaotic', is \emph{inherently} unpredictable---not merely practically or due to the butterfly effect. The radical implications of this realization will be discussed elsewhere. }

\subsection{The `observer' in the block universe}\label{observer}
The term `observer' has permeated physics texts only with the birth of QM. It's not that there were no Newtonian or even Einsteinian observers, making measurements from various angles, it's just that physicists back then focused their attention on something they believed lied outside their personal experience; something objectively `out there'.

So what exactly happened in the 1920's that changed this view? One factor was Heisenberg's realization that some aspects of a system cannot be measured without being perturbed. This is, of course, true also in classical physics, but in that case,  the perturbation can---at least in principle---be  made arbitrarily small. If something cannot be measured, it was argued, then in what sense does it objectively exist?

The second, stronger reason for introducing the observer right into the \emph{formalism} of QM is mathematical: Unlike classical mechanics, the Schr{\"o}dinger equation cannot possibly describe some objective reality waiting  to be measured. It is fundamentally a statistical description of measurements---unitarity built right into it for that cause---inexorably demanding an observer. 

Over the years, many attempts have been made to do away with the quantum observer, thereby restoring a (somewhat) classical description of physics,  but the block-universe, as we show next, undoubtedly offers the most direct way of doing so.

Imagine, then, an ECD block-universe, in which particles  emit, absorb and scatter EM waves. The block-universe, being as big as it is---some space-time regions in it contain e-m distributions which undoubtedly resemble those appearing in laboratories conducting QM experiments.
For concreteness, let us look at a space-time region in which the following  scenario takes place: An atom sitting in a dark cave is excited by an energetic cosmic particle, and then spontaneously decays, emitting EM radiation which is scattered by some crystal (who has acquired its orientation through passive growth over millions of years, unaware of its future role as a beam-splitter...) and ultimately triggers an irreversible macroscopic process, leaving a visible mark on the wall of the cave. Much later, an  `observer'  arrives at the cave and, like a good forensic scientist, he is able to reconstruct a reasonable approximation of the relevant e-m (space-time) distribution representing the above scenario. Our observer is essentially the same passive classical observer, registering the orbits of planets, and after successive  visits to the cave, and to other caves, he can compute the relative frequency of the various e-m distributions\footnote{As two successive decays necessarily correspond to two e-m distributions which are supported on different times, the distributions are defined modulo a time-shift.}  which, so he believes like any scientist, faithfully represent the frequency of their  occurrence in the entire block-universe.

\begin{figure}
	\centering
		\includegraphics[width=.800\textwidth]{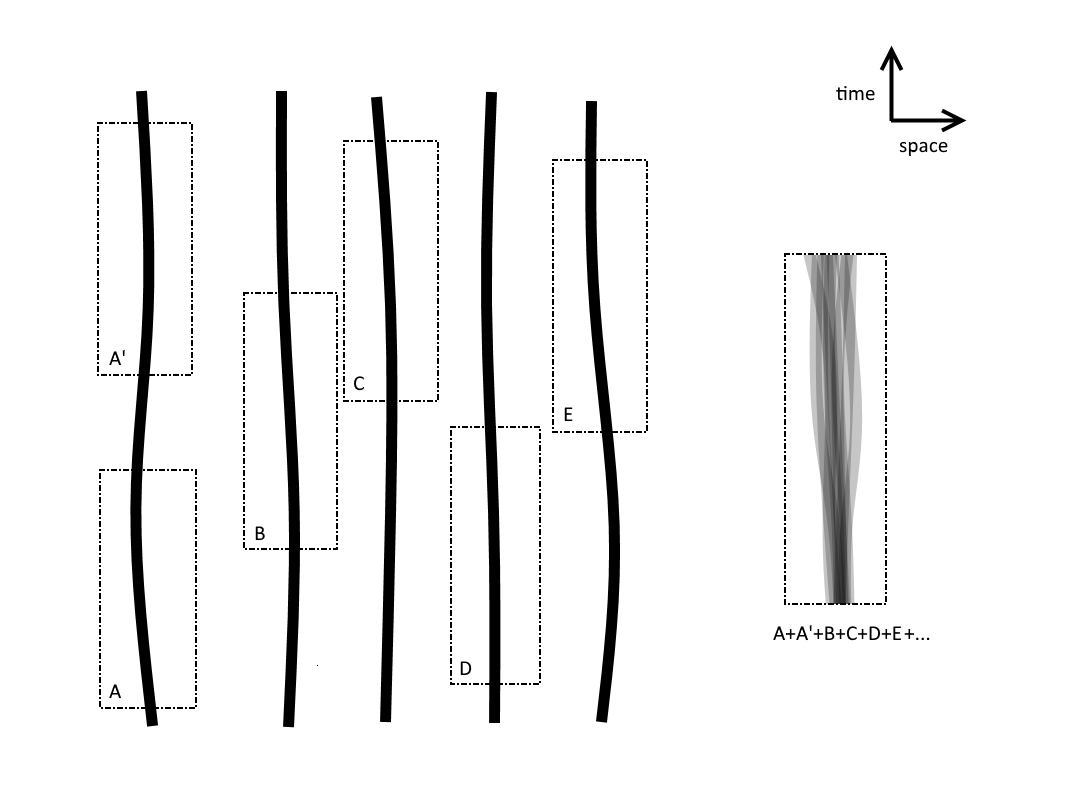}
	\caption{\footnotesize{Typical time slices (A,A',B,C,D,E) taken from the e-m/charge distribution associated with individual systems. Slices A, A', A''... represent, e.g., a single hydrogen atom at different times, whereas B,C,D... represent multiple atoms involved in a scattering experiment.  By `superimposing' all members of the ensemble (right), one gets various ensemble densities, e.g. charge density $\rho_\text{ens}$, from which the (statistical) results of any QM experiment can be deduced. From the basic tenets constraining individual members in the ensemble, equations for the ensemble densities can be derived and, upon writing $\rho_\text{ens}\equiv\phi^\dag\phi$, QM wave equations for $\phi$ follow. The different ways of superimposing the different members, correspond to different solutions of the same equation. A straightforward generalization to non relativistic many-body non-machines also exists.  }  }
	\label{fig:spacetime}
\end{figure}

Equipped with his relative frequencies data, the observer then arrives at a new cave only to discover that an atomic decay has not yet occurred. He therefore cannot, yet, decide which e-m distribution in the data he previously collected, corresponds to the about-to-happen decay, but he still has some partial information, such as the orientations of the beam-splitters, the type of atom, etc., so he can focus only on that subset of collected e-m distributions which are consistent with his partial knowledge, and compute the probabilities of obtaining, say, a mark on the wall at a certain position. This sort of statistical inference is QM, seen from the perspective of a `BU observer' (see figure \ref{fig:spacetime}).

In practice, searching for deserted caves for traces of spontaneous decays of atoms, is expensive. Experimental data is therefore collected in a single cave (lab) instead, with Ph.D students or, equivalently, robots with some RNG component in their software, determining the orientation of the beam-splitters.  But should that matter? Should we expect that the statistical regularities prevailing in the rest of the universe would not apply to the laboratory?

\section{The meaning of the wave function}\label{The ensemble current}
Our block-universe is allegedly described by a specific global ECD solution whose exact form will forever remain concealed. Nevertheless, in many cases, it is not the exact solution which is of interest but rather certain statistical properties of it. Guided by the basic tenets, locally satisfied throughout the entire ECD solution, we derive in this section   a plausible statistical description of the ECD block-universe.
We show that QM is such a theory, and not by a pure guess:  Like ECD, QM is  a realization of the basic tenets, only with two less constraints:  Being a statistical theory of \emph{ensembles} of particles with a \emph{common} mass, it does not have to describe localized currents and it should not be scale covariant. It is these two added constraints which make it so difficult to solve the classical self-force problem, possibly explaining why QM could have pre-dated ECD's realization in \cite{YK2016} (assuming, of course, the latter is a valid theory...).

Let us, then, look at a collection of time slices of the ECD block-universe, viz. segments of the block-universe bounded by two space-like surfaces of a given separation in time,  each corresponding to a repetition of an experiment. The e-m distribution in each such time-slice may involve a different distinguished particle, as in a scattering experiment, or the same particle---say, a radiating electron in a trap---at different times.  If we now shift all time slices to a common support in time, and add the distinguished current in each, we get for our distinguished particle an electric  \emph{ensemble current}. We shall later give specific examples of ensemble currents appearing in various QM contexts, but the reader should easily convince himself, already at this stage, that the results of any single particle QM experiment  can be read from the ensemble current---an  ordinary, conserved four-current. Just as an example: Far away from a scatterer, the intensity of the ensemble current associated with a scattering experiment, measures the likelihood of a particle to scatter to a given angle.  

Consider, next, a time-slice in the ensemble, indexed by $e$. We decompose the global EM potential,  \eqref{convolution with K}, in that slice  into an external piece, $A_\text{ext}$,  assumed constant throughout the ensemble, which is generated by a likewise $e$-independent $j_\text{ext}$, and a fluctuating piece, $\lsuper{e}\tilde{A}$,  which varies across the ensemble, incorporating  the $e$-specific field generated by the   electric current, $\lsuper{e}j$, associated with the distinguished particle. The fluctuating part, satisfying  $\partial\, \lsuper{e}\tilde{F}=\lsuper{e}j$, is obviously  defined only up to a  homogeneous solution of Maxwell's equation, and therefore incorporates also a (possible) radiation field generated by the rest of the particles in the universe (see section \ref{conspiracy} for details).

Thus to each time-slice, $e$, in the ensemble  correspond   a distinguished   electric current $\lsuper{e}j$, an e-m tensor    $\lsuper{e}T$,  an EM potential $\lsuper{e}\tilde{A}$ which, in the vicinity of our distinguished particle, satisfy the basic tenet \eqref{ni} 
\beq\label{ghgh}
\partial_\nu \lsuper{e}T^{\nu \mu} =\left(F_\text{ext}{}^{\mu \nu} + \lsuper{e}\tilde{F}^{\mu \nu}\right) \;\lsuper{e}j_\nu\equiv \lsuper{e}f_\text{ext}{}^\mu + \lsuper{e}f_\text{sel}{}^\mu\,,
\enq
with $ f_\text{ext}$ and  $f_\text{sel}$ the external and self forces respectively,
and similarly for the e-m tensor associated with the external source
\beq\label{ghgh-ext}
\partial_\nu T_\text{ext}{}^{\nu \mu} =\left(F_\text{ext}{}^{\mu \nu} + \lsuper{e}\tilde{F}^{\mu \nu}\right) j_\text{ext}{}_\nu\,,
\enq
which, like $j_\text{ext}$, is assumed $e$-independent as a first approximation.
Applying Poynting's theorem \eqref{Poynting}  to the global field $F_\text{ext}+\lsuper{e}\tilde{F}$  we get  
\beq\label{ghgh2}
\partial_\nu \lsuper{e}\Theta^{\nu \mu} =-F_\text{ext}{}^\mu_{\phantom{1} \nu}\,j_\text{ext}{}^\nu-F_\text{ext}{}^\mu_{\phantom{1} \nu}\,\lsuper{e}j^\nu -\lsuper{e}\tilde{F}^\mu_{\phantom{1} \nu}\,j_\text{ext}{}^\nu-f_\text{sel}{}^\mu\,.
\enq 
With the above definitions, local e-m conservation \eqref{pp} of the system comprising the distinguished particle and the source is guaranteed even when those two overlap, e.g., in a Hydrogen atom.

If we now assume that each member in the ensemble current is (fairly) sampled from a dense set of currents, indexed by $e$, with normalized measure $\mu$ ($\int\d\mu(e)=1$), we can integrate \eqref{ghgh} over $e$ with respect to $\d\mu(e)$ (removing possible integrable singularities in the individual currents as a by-product). Defining the following ensemble quantities for our distinguished particle
\begin{align}\label{summ}
j_\text{ens}=\int\d \mu(e)\,\lsuper{e}j&\,,\quad T_\text{ens}=\int\d \mu(e)\,\lsuper{e}T\,,\quad f_\text{ens}{}^\mu=\int\d \mu(e)\,\lsuper{e}\tilde{F}^{\mu \nu} \;\lsuper{e}j_\nu\\
& \Theta_\text{ens}=\int\d \mu(e)\,\lsuper{e}\Theta\,\quad  \tilde{F}_\text{ens}=\int\d \mu(e)\,\lsuper{e}\tilde{F}\,,\nonumber
\end{align}
we obtain smooth quantities satisfying the basic tenets \eqref{ni} and \eqref{Maxwell's_equation} respectively,
\begin {equation}\label{hcc}
 \partial_\nu T_\text{ens}{}^{\nu \mu}=F_\text{ext}{}^\mu_{\phantom{1} \nu}\,j_\text{ens}{}^\nu + f_\text{ens}{}^\mu\,,
\end{equation}
\beq\label{ensemble_field}
\partial_\nu \tilde{F}_\text{ens}{}^{\nu\mu}=j_\text{ens}{}^\mu\,.
\enq 
To add $\Theta_\text{ens}$ to the discussion, it is first convenient to decompose it into three different parts
\begin{align}\label{Theta_defined}
&\Theta_\text{ext}{}^{\mu\nu}=\frac{1}{4} g^{\nu \mu} F_\text{ext}{}^{\rho\lambda}F_\text{ext}{}_{\rho\lambda} + F_\text{ext}{}^{\nu \rho}F_\text{ext}{}_\rho^{\phantom{1} \mu}\\
&\Theta_\text{int}{}^{\mu\nu}=\half g^{\mu\nu} \tilde{F}_\text{ens}{}^{\rho\lambda} F_\text{ext}{}_{\rho\lambda}+ \tilde{F}_\text{ens}{}^{\mu\lambda} F_\text{ext}{}_\lambda^{\phantom{1}\nu} + F_\text{ext}{}^{\mu\lambda} \tilde{F}_\text{ens}{}_\lambda^{\phantom{1}\nu}\,\nonumber\\
&\Theta_\text{sel}{}^{\mu\nu}=\int\d\mu(e)\frac{1}{4} g^{\nu \mu}\, \lsuper{e}\tilde{F}^{\rho\lambda}\,\lsuper{e}\tilde{F}_{\rho\lambda} + \lsuper{e}\tilde{F}^{\nu \rho}\,\lsuper{e}\tilde{F}_\rho^{\phantom{1} \mu}\,,\nonumber
\end{align}
which, using \eqref{ghgh2}, satisfy
\begin{align}\label{hcc2}
&\partial_\nu \Theta_\text{ext}{}^{\nu \mu}=-F_\text{ext}{}^\mu_{\phantom{1} \nu}\,j_\text{ext}{}^\nu\,,\qquad \partial_\nu \Theta_\text{sel}{}^{\nu \mu}=-f_\text{ens}{}^\mu\,,\\
&\partial_\nu \Theta_\text{int}{}^{\nu \mu}=-F_\text{ext}{}^\mu_{\phantom{1} \nu}\,j_\text{ens}{}^\nu -\tilde{F}_\text{ens}{}^\mu_{\phantom{1} \nu}\,j_\text{ext}{}^\nu\,.\nonumber
\end{align}

The ensemble quantities  are, in general,  much duller objects than the individual quantities composing them. As we shall see in section \ref{conspiracy}, those latter are rather chaotic, but their chaotic components superpose incoherently  and vanish following the integral over $e$.
Note also that it is the measure, $\d\mu$, representing the frequency of occurrence  of certain e-m configurations in the global ECD solution, which encodes the results of QM experiments. And since it is such an abstract object, we human cannot judge it purely on the basis of our intuition.

Next, we make the following approximation whose validity is critical to the rest of our analysis: We neglect the $f_\text{ens}(x)$. From  \eqref{hcc}, \eqref{hcc2}, and $f_\text{ens}=0$  we immediately get that the combined tensor,
$T_\text{ext}+\Theta_\text{ext}+T_\text{ens}+\Theta_\text{int}+\Theta_\text{sel}$, is locally conserved in regions of space-time containing our distinguished and source particles. Since  $f_\text{ens}=0$ implies $\partial_\nu\Theta_\text{sel}^{\nu\mu}= 0$, we can further `absorb' $\Theta_\text{sel}$ into the definition $T_\text{ens}$, consistently allowing for the latter to incorporate also the  Coulomb self energy of members in the ensemble (see appendix D of \cite{YK2016}).

To see when our approximation is justifiable, let us write  the  self force on an individual ensemble member as $\lsuper{e}f_\text{sel}=f_\text{sys}+\lsuper{e}\tilde{f}$, where $f_\text{sys}$ is a   `systematic' part, common to all members, and $\lsuper{e}\tilde{f}$ is a member-specific fluctuating piece. As an example for such a decomposition, consider a freely moving charge. In its rest frame, $f_\text{sys}$ is likely to be dominated by a  spherically symmetric outward pointing radial filed, coming from the near, viz,. Coulomb, self field, with a support coinciding with that of $\lsuper{e}j^0$, and  $\lsuper{e}\tilde{f}$ is at most a relatively small, randomly pointing field which can safely be assumed to enter incoherently into $f_\text{ens}$ and is therefore ignored. The systematic part, in contrast, can only be ignored if $j_\text{ens}{}^0$ varies slowly over the extent of $\lsuper{e}j^0$. More precisely, in the above example the ensemble self-force takes the form of a convolution 
\beq\label{convolution}
{\boldsymbol f}_\text{ens}({\boldsymbol x})=\int\d^3x'\,K({\boldsymbol x}'){\boldsymbol f}_\text{sys}({\boldsymbol x}-{\boldsymbol x}')\,,
\enq 
which, for spherically symmetric ${\boldsymbol f}_\text{sys}$, can only be assumed small if $K$ is slowly varying over the support of ${\boldsymbol f}_\text{sys}$. 

Another consistency condition regarding the omission of $f_\text{ens}$ concerns the spherical symmetry of  ${\boldsymbol f}_\text{sys}$.  Any external field, necessarily breaking the isotropy of space, is expected to violate our assumption to some degree. In particular, the magnetic self-force which in the previous free particle case was `pushed' into the $\lsuper{e}\tilde{f}$, randomly oriented part of the self force, should now contribute to ${\boldsymbol f}_\text{sys}$ as well. 

Summarizing, our treatment of the self-force is clearly only an approximate one, expected to break down when strong, or rapidly varying  external fields are involved.

\subsection{Relativistic single particle QM wave equations}

Although Schr{\"o}dinger did not even mention the basic tenets in his first attempt to write a `wave equation for matter', his proposal, latter dubbed the Klein-Gordon (K-G) equation, provides  a  systematic way of obtaining  a conserved current $j_\text{ens}$ which, along with $T_\text{ens}$, satisfies the $f_\text{ens}$-free \eqref{hcc}. Starting with solutions to the K-G equation
\beq\label{KG}
\left(D^\mu D_\mu +m^2\right)\phi=0\,,
\enq
where  $D = \hbar\partial - i q A_\text{ext}$ is the covariant derivative, $\hbar,q$ and $m$ are arbitrary  parameters (and $c=1$), we get our desired conserved electric current 
\beq\label{KG_current}
 j^\mu=q\,\text{Im }\phi^* D^\mu\, \phi\,,
\enq
and  symmetric e-m tensor
\beq\label{ksh}
 T^{\nu\mu}=\frac{g^{\nu \mu}}{2} \left(   m^2\phi^*\phi -
  \left(D^\lambda \phi\right)^* D_\lambda \phi \right) + \frac{1}{2}\big(D^\nu
               \phi \left(D^\mu \phi\right)^* + \text{c.c.}
 \big)\,.
\enq

%

The fact that $j^0_\text{ens}$ above can take both positive and negative values, which has troubled the founders of QM, can have two, not mutually exclusive meanings in our approach. The first is that it corresponds to an ensemble of charged particles and their antiparticles. As a consistency check we verify that $\phi(-x)$ solves  \eqref{KG} for $-A(-x)$, and we get another ensemble, in a potential represented by $-A(-x)$, in which every particle in the original ensemble would be replaced by its antiparticle,  consistent with symmetry \eqref{PT}. A second interpretation, however, is that the current \eqref{KG_current} represents an ensemble of charged particles, each with a charge distribution  taking both signs---a `more interesting' sort of particle. 

Another covariant way of obtaining our desired enesemble quantities \eqref{summ} begins with Dirac's equation (see, e.g. \cite{Itzykson} for notations)
\beq\label{Dirac_equation}
i\gamma^\mu D_\mu\psi=m\psi\,.
\enq
The conserved electric current associated with spinor solutions $\psi$
\beq\label{Dirac_current}
j^\mu=q\bar{\psi}\gamma^\mu\psi
\enq
has a charge density, $j^0$, with a sign which everywhere equals to $\text{sign}(q)$ (unlike in the K-G case). To be compatible with the normalization of the measure $\mu$, and the total charge, $q$, of individual particles, the usual normalization $\int\d^3x\,\psi^\dagger\psi=1$ must be imposed on $\psi$.  Along with   
\beq\label{Dirac_T}
T^{\mu\nu}=\frac{i}{4}\left( \bar{\psi}\gamma^\mu \overleftrightarrow{D}^\nu\psi +  \bar{\psi}\gamma^\nu \overleftrightarrow{D}^\mu\psi \right)-\frac{g^{\mu\nu}}{2}\left( i\bar{\psi}\gamma^\lambda \overleftrightarrow{D}_\lambda\psi - 2m\bar{\psi}\psi\right)\,,
\enq
the $f_\text{ens}$-free \eqref{hcc} is satisfied. 

Both the Dirac and KG realizations of the basic tenets involve three tunable parameters: $q$, $m$ and $\hbar$, the meaning of the last is discussed bellow.  When, as in the Dirac case,  $q$ is the charge of the particle, then by Ehrenfest's theorem (applied to a localized wave-packet moving in a weak, slowly varying $F_\text{ext}$) independently of $\hbar$, $m$ appears as the classical mass of particle which, by our previous remark, incorporates also the Coulomb self energy.

Contrary to the K-G case, in the rest frame of a wave packet solution of Dirac's equation,  a non-vanishing angular momentum,  $J_\imath=\int\d^3x\,\epsilon_{\imath\jmath k} x_\jmath T^{0k}$, and  magnetic dipole moment $\mu_\imath=\half\int\d^3x\,\epsilon_{\imath\jmath k} x^\jmath j^k$, appear, consistent with an underlying ensemble of particles, each with an internal current---so-called `spinning particles'. Note, nonetheless, that both equations are realizations of the same basic tenets. The spinorial nature of $\psi$ (as does the complex nature of $\phi$) merely reflects the nature of the `scaffold' used to construct real tensors, transforming under integer representations of the Lorentz group. 

Remarkably, if the external field is purely electrostatic, and the Coulomb gauge is being used,  the energy density associated with the distinguished particle, $T_\text{ens}{}^{00}+\Theta_\text{int}{}^{00}$, with $T_\text{ens}$ given by \eqref{Dirac_T} and $\Theta_\text{int}$ defined in \eqref{Theta_defined}, equals $\text{Re}\, \bar{\psi}{\cal H}\psi $, with 
\beq\label{Hamiltonian}
{\cal H}=-i\gamma^0\gamma^iD_i + \gamma^0 m + q\gamma^0A^0
\enq
the usual Hamiltonian governing the evolution of $\psi$, i.e. $i\hbar\partial_t\psi={\cal H}\psi$, and we note that, by the Hermiticity of ${\cal H}$, the imaginary part of  $\bar{\psi}{\cal H}\psi$ anyway does not contribute to its `expectation value' (its  integral over three space). Note that this standard energy density form, in contrast to our proposal, is \emph{not} gauge invariant which is unaccepted from a physical standpoint. The  breaking of both  manifest Lorentz and gauge covariance of Dirac's equation, resulting from the use of the Hamiltonian formalism, has (to best of the author's knowledge) never found a satisfactory solution (see e.g. \cite{Reiss} ). By using our alternative e-m tensor, a manifestly covariant (Dirac) equation, and an ensemble interpretation, gauge invariant results are automatically obtained.

\subsubsection{The $\hbar\rightarrow 0$ limit}\label{hbar limit}
As we shall see in the next section, the self-force  plays a crucial role in the description of individual members in the ensemble (in particular, the very existence of a particle involves the self-force. See section 3.2 in \cite{YK2016}). The elimination of $f_\text{ens}$ from the equations for the ensemble currents only means that the statistical effects associated with the self-force can be absorbed into the free parameters of the wave-equations and in this section  we argue that for a fixed $m$ and $q$, it is   $\hbar$ which, roughly speaking, measures the strength of self-force effects.

A direct demonstration of this claim starts with the trajectory of a point charge $\gamma^\mu(\tau)$ ($\tau$ the proper time)  in an external field $F_\text{ext}$, which satisfies the Lorentz force equation $m\ddot\gamma=qF(\gamma)\dot{\gamma}$. Associated with it is an e-m tensor,  
\beq\label{classical_m}
T^{\nu\mu}=m\int_{-\infty}^{\infty}\d \tau\,\,
 \dot{\gamma}^\nu \, \dot{\gamma}^\mu\,
 \delta^{(4)}\big(x-\gamma\big)\, ,
\enq
and a conserved electric current
\beq\label{classical_j}
j^{\mu}=q\int_{-\infty}^{\infty}\d \tau\,\,
 \dot{\gamma}^\mu 
 \delta^{(4)}\big(x-\gamma\big)\, ,
\enq
satisfying (in the distributional sense) the $f_\text{ens}$-free \eqref{hcc}
\begin{align}
\partial_\nu T^{\nu \mu}&=m\int\d \tau\,\,
 \dot{\gamma}^\nu \, \dot{\gamma}^\mu\,
 \partial_\nu\delta^{(4)}\big(x-\gamma\big)=-m\int\d \tau\,\,
  \dot{\gamma}^\mu\,
 \partial_\tau\delta^{(4)}\big(x-\gamma\big)\nonumber\\
 &=\int\d \tau\,\, m\ddot{\gamma}^\mu\,\delta^{(4)}\big(x-\gamma\big)=\int\d \tau\,\,qF_\text{ext}^{\mu\nu}\dot{\gamma}_\nu\,\delta^{(4)}\big(x-\gamma\big)=F_\text{ext}^{\mu\nu}\,j_\nu\,.\nonumber
\end{align}
By linearity, therefore, any ensemble of charges, with associated $j_\text{ens}$ and $T_\text{ens}$ of the form appearing in \eqref{summ}, which uses the above distributions for $\lsuper{e}j$ and $\lsuper{e}T$, also satisfy \eqref{hcc}.  The direct method of demonstrating the above stated role of $\hbar$ relies on the following result which is a degenerated version of the Bohm-De Broglie theory: In the limit $\hbar\rightarrow 0$, to every electric current, $j$, and e-m tensor, $T$,  generated  from a wave-function solution, of either the Dirac or K-G equation,  correspond $j_\text{ens}$ and $T_\text{ens}$ generated by the above procedure for an appropriate measure $\mu$. In other words, Dirac and K-G densities appear in the $\hbar\rightarrow 0$ limit as if composed of point particles, moving in an external field.  

But the above result suggests two additional things: Not only self-force corrections to the Lorentz force equation disappear in the limit $\hbar\rightarrow 0$. Spin effects likewise disappear, along with the size of the particle which shrinks to a point (this last consequence is further discussed in section \ref{Compton}).

\subsection{The many-body wave function}\label{MBWF}
In experiments involving two or more particles, identical or otherwise, the individual ensemble currents associated with each particle tell us a very partial story. They do not, for example, tell us anything about the results of correlation experiments between the spins of two particles, of the type used in Bell's inequality tests.   We therefore seek a richer mathematical object, capable of describing such correlations.

Regrettably, a simple generalization of the construction leading to the single-body ensemble current runs into the problem that each member of the many-body ensemble involves a different EM field which, unlike the self-field, cannot (and should not) be eliminated from the ensemble current. However, at small particle velocities (compared with the speed of light) a great simplification arises by virtue of the fact that the  inter-particle interaction is well approximated by the Coulomb interaction, instantaneous in some reference frame, suggesting a nonrelativistic construction. It is emphasized, nonetheless, that by treating time as a distinguished coordinate, \emph{in no way do we  abandon the block-universe view}. We are still attempting to describe the statistics of space-time structures, only this time we choose to view them as a union of `space-like slices' indexed by time in some convenient coordinate system.

Anticipating the universal applicability of Schr{\"o}dinger's equation, treating  manifestly composite particles, such as atoms,  molecules, or even  macroscopic BEC's (e.g. q-bits) as  indivisible entities, we begin by grouping together elementary particles forming a composite, denoting by $C_a$ the subset of particles forming composite $a$, and by $j^{(a)}=\sum_{a'\in C_a}j^{(a')}$  and  $T^{(a)}=\sum_{a'\in C_a}T^{(a')}$ their collective current and e-m tensors resp., and note that the basic tenets retain their form when the particle index refers to composites rather than elementary particles. 

We proceed by recasting the basic tenets into a non-relativistic form, in which instantaneous action-at-a-distance makes sense. To this end we first define a normalized charge density $\rho^{(a)}\equiv j^{(a)\,0}/q^{(a)}$ and normalize the electric current  $j^{(a)\,i}\mapsto j^{(a)\,i}/q^{(a)}$, with $q^{(a)}=\int\d^3x\,j^{(a)\,0}$ the conserved charge. By \eqref{continuity_equation}, these satisfy 
\beq\label{mc}
 \partial_t\rho^{(a)}=-\partial_ij^{(a)\,i}\,.
\enq

Next, we integrate the basic tenet  \eqref{ni} over three space, treating the $\mu=1,2,3$ case first. As in the single particle case, we decompose the EM field on the r.h.s. of \eqref{ni} into an external piece, and a self field. The first is composed of external electric and magnetic fields,
\beq\label{potentials}
{\boldsymbol E}=-{\boldsymbol \nabla}\varphi-\partial_t{\boldsymbol A}\,,\quad {\boldsymbol B}={\boldsymbol \nabla}\times {\boldsymbol A}\,,
\enq
with potentials assumed to be given in the (instantaneous) Coulomb gauge,
plus the EM field generated by all \emph{other} particles in the group. The electric component of that field is given in the Coulomb gauge by  $-{\boldsymbol \nabla} V^{(a)}-\partial_t {\boldsymbol A}^{(a)}$, with 
\beq\label{Laplace}
\nabla^2V^{(a)}({\boldsymbol x},t)=-\sum_{b\neq a} q^{(b)}\rho^{(b)}({\boldsymbol x},t)\,,\quad \lim_{|{\boldsymbol x}|\rightarrow \infty} V^{(a)}({\boldsymbol x},t)=0\,.
\enq 
By our assumption of slowly moving particles,  the corresponding expression for the vector potential, ${\boldsymbol A}^{(a)}$, can be shown to be negligibly small in the Coulomb gauge.  However, we shall be considering also so-called `spinning particles'---particles with an internal current---which, even at rest, generate a magnetic field. These are also neglected, the justification being this time their  smallness, and we shall have to validate the consistency  of this approximation in the sequel. 
  
Returning to the integration of  \eqref{ni}, the divergence $\partial_iT^{(a)\,i\mu}$ drops due to the localization of $T^{(a)}$. Defining  $\langle\cdot\rangle\equiv \int\d^3 x\, \cdot\,$  we therefore get the Lorentz force equation 
\begin{align}\label{P_i}
\frac{\d}{\d t} \left\langle p^{(a)}_i\right\rangle= &-q^{(a)}\Big\langle\partial_i V^{(a)}\rho^{(a)}\Big\rangle+q^{(a)}\Big\langle E_i\rho^{(a)}\Big\rangle\\& +q^{(a)}\left\langle \epsilon_{ikl}j^{(a)\,k} B_l\right\rangle   + f_\text{sel\,i}^{(a)}(t)\qquad\text{(no summation on $a$)}\,, \nonumber
\end{align}
with $p^{(a)}_i= T^{(a)\,0i}$ the $i$th component of the three momentum density, $j$ the normalized current, and $f_\text{sel\,i}^{(a)}(t)$ the self force which is already a result of a space integral, hence the dominant radial self Coulomb force on a spherically symmetric $\rho$ is eliminated already at this stage.

Repeating the procedure for the $\mu=0$ component of \eqref{ni}, we get
\beq\label{energy_density}
\frac{\d}{\d t}\Big\langle\varepsilon^{(a)}\Big\rangle=q^{(a)}\Big\langle \left(E_i-\partial_iV^{(a)}\right) j^{(a)\,i}\Big\rangle+f_\text{sel\,0}^{(a)}\,,
\enq
with $\varepsilon^{(a)}=T^{(a)\,00}$ the  energy density associated with a particle.

Next, we integrate the $\mu=0$ component of Poynting's theorem \eqref{Poynting} over a volume containing our particles. The resulting r.h.s. is just minus the r.h.s. of \eqref{energy_density}, summed over $a$.  Summing \eqref{energy_density} over $a$ and adding the two, we therefore get
\beq\label{first_energy}
\frac{\d}{\d t}\left\langle\sum_a\varepsilon^{(a)}+\Theta^{00}\right\rangle=-\int_S \d {\boldsymbol S}\cdot{\boldsymbol P}\;,
\enq
where $S$ is a surface enclosing our particles and ${\boldsymbol P}={\boldsymbol E}_\text{total}\times{\boldsymbol B}_\text{total}$ is the  Poynting flux, possibly incorporating   also  the field radiate by our particles. Note that for the l.h.s. of \eqref{first_energy} to include only the $\varepsilon$'s of our interacting particles, we must assume that the particles generating the external field lie outside $S$, or else that their energy is approximately time-independent (as is the case with the proton's kinetic energy in he electron-proton Hydrogen system).

As in the single particle case, $\Theta^{00}=\half\left( {\boldsymbol E}_\text{total}^2+{\boldsymbol B}_\text{total}^2\right)$ can be written as a sum: $\Theta^{00}=\Theta_\text{ext}{}^{00}+\sum_a\Theta_\text{sel}^{(a)}{}^{00}+\Theta_\text{int}{}^{00}$. The sel(f) parts are dominated by a quadratic term  in the self Coulomb field, contributing approximately a time independent constant (unless a particle suffers a severe deformation, which we assume not to be the case).  The remaining contributions to $\Theta^{00}$ are a trivial ext(ernal) piece, composed of quadratic terms in the external EM field, and an int(eraction) piece, coming from bilinears in the external and self fields. Using Maxwell's equations and the Coulomb gauge condition, ${\boldsymbol \nabla}\cdot{\boldsymbol A}=0$, the integral of the int(eracttion) piece can be shown to equal 
\beq\label{hvv}
\left\langle\Theta_\text{int}{}^{00}\right\rangle= \sum_a q^{(a)}\left\langle{\boldsymbol j}^{(a)}\cdot{\boldsymbol A} + \left( \varphi + V^{(a)}\right)\rho^{(a)} \right\rangle\,.  
\enq
Note that since our particles are assumed to be moving slowly, the normalized $j$ appearing above is (at most) due to a current internal to a particle. By our assumption that this current generates a small magnetic field, we can, as with the self magnetic energy, also neglect bilinears in this field appearing in $\Theta_\text{int}$ (each coming from a different particle).

To convert the above  relations into equations for the ensemble quantities, we first create the following multi-particle counterparts of the single-particle densities
\beq\label{dnn}
\lsuper{e}\rho\left({\boldsymbol x}^{(1)},\ldots,{\boldsymbol x}^{(n)},t\right) =\prod_a \,\lsuper{e}\rho^{(a)}\left({\boldsymbol x}^{(a)},t\right)\,,
\enq
\beq\label{jnn}
\lsuper{e}j_{i_a}\left({\boldsymbol x}^{(1)},\ldots,{\boldsymbol x}^{(n)},t\right) = \lsuper{e}j^{(a)\,i}\left({\boldsymbol x}^{(a)},t\right)\prod_{b\neq a}\,\lsuper{e} \rho^{(b)}\left({\boldsymbol x}^{(b)},t\right)\,,
\enq 
\beq\label{pnn}
\lsuper{e}p_{i_a}\left({\boldsymbol x}^{(1)},\ldots,{\boldsymbol x}^{(n)},t\right) = \lsuper{e}p^{(a)}_i\left({\boldsymbol x}^{(a)},t\right)\prod_{b\neq a}\,\lsuper{e} \rho^{(b)}\left({\boldsymbol x}^{(b)},t\right)\,,
\enq 
\beq\label{En}
\lsuper{e}\varepsilon_a\left({\boldsymbol x}^{(1)},\ldots,{\boldsymbol x}^{(n)},t\right) = \lsuper{e}\varepsilon^{(a)}\left({\boldsymbol x}^{(a)},t\right)\prod_{b\neq a}\,\lsuper{e} \rho^{(b)}\left({\boldsymbol x}^{(b)},t\right)\,,
\enq
where we have restored the time-slice index, $e$, omitted  in this section.
Defining $\left\langle\cdot\right\rangle_{{\mathbb R}^{3n}}\equiv\int\prod_{a}\d^3 x^{(a)}\,\cdot\quad$ , and $\partial_{i_a}\equiv \partial/\partial x^{(a)}_i$, \eqref{mc}, \eqref{dnn} and \eqref{jnn} imply a $3n$-dimensional continuity equation
\beq\label{mpCont}
\partial_t\,\lsuper{e}\rho=-\sum_{i,a}\partial_{i_a}\,\lsuper{e}j_{i_a}
\enq
Similarly, \eqref{P_i} and \eqref{energy_density} give, respectively
\begin{align}\label{mpNewton}
\frac{\d}{\d t} \Big\langle \lsuper{e}p_{i_a}\Big\rangle_{{\mathbb R}^{3n}} =& -\Big\langle\partial_{i_a} V\,\lsuper{e}\rho\Big\rangle_{{\mathbb R}^{3n}} +q^{(a)}\Big\langle E_i\left({\boldsymbol x}^{(a)},t\right)\, \lsuper{e}\rho\Big\rangle_{{\mathbb R}^{3n}}\\& + q^{(a)}\Big\langle \epsilon_{ikl}\,\lsuper{e}j_{k_a}B_l\left( {\boldsymbol x}^{(a)},t\right) \Big\rangle_{{\mathbb R}^{3n}} +\lsuper{e}f_{\text{sel}\,i}^{(a)}(t)\nonumber\,,
\end{align}
\beq\label{mp_energy_density}
\frac{\d}{\d t}\Big\langle\,\lsuper{e}\varepsilon_a\Big\rangle_{{\mathbb R}^{3n}}=q^{(a)}\Big\langle E_i\left({\boldsymbol x}^{(a)},t\right) \lsuper{e}j_{i_a}\Big\rangle_{{\mathbb R}^{3n}}
-q^{(a)}\Big\langle \partial_{i_a}V\,\lsuper{e} j_{i_a}\Big\rangle_{{\mathbb R}^{3n}}
+\lsuper{e}f_\text{sel\,0}^{(a)}\,,
\enq
\beq\label{mppotential}
V=\frac{1}{8\pi}\sum_{{}^{b, a}_{a\neq b}} \frac{q^{(a)}q^{(b)}}{\left|{\boldsymbol x}^{(a)} - {\boldsymbol x}^{(b)}\right|}\,, 
\enq 
where \eqref{mpNewton} and \eqref{mp_energy_density} follow from the Green's function of the Laplacian
\beq 
\nabla^2\,\frac{-1}{4\pi|{\boldsymbol y}-{\boldsymbol y'}|}= \delta^{(3)}({\boldsymbol y}-{\boldsymbol y'})\,.\nonumber
\enq
Finally, the interaction energy \eqref{hvv} associated with time slice $e$, takes the form
\beq\label{interaction_energy}
\Big\langle\lsuper{e}\Theta_\text{int}{}^{00}\Big\rangle=\Big\langle V\,\lsuper{e}\rho\Big\rangle_{{\mathbb R}^{3n}}+\sum_a q^{(a)}\Big\langle \lsuper{e}j_{i_a}A_i\left( {\boldsymbol x}^{(a)},t\right)\Big\rangle_{{\mathbb R}^{3n}}+ q^{(a)}\Big\langle \lsuper{e}\rho\varphi\left( {\boldsymbol x}^{(a)},t\right)\Big\rangle_{{\mathbb R}^{3n}}
\enq

We are now in a position to look at ensemble quantities. Repeating the steps leading to the ensemble currents in the single-particle case, we  multiply each equation by a normalized measure, $\d\mu(e)$,  giving $e$'s `probability density' to be realized in an experiment, and integrate over $e$. The reader can verify that the results of every conceivable multi-particle nonrelativistic  QM experiment is encoded in the four ensemble densities $d_\text{ens}\equiv \int\d\mu(e)\, \lsuper{e}d$, $d=\rho,j,p,\varepsilon$. For further reference we shall also associate with each, $n$  \emph{marginal densities}
\begin{align}\label{marginal_densities}
d^{(a)}_\text{mar}({\boldsymbol x}^{(a)},t)&:=\int\Pi_{b\neq a}\d^3 x^{(b)}\, d_\text{ens}=\int\d\mu(e) \int\Pi_{b\neq a}\d^3 x^{(b)}\, \lsuper{e}d\\
&=\int\d\mu(e) \, \lsuper{e}d^{(a)}\,,\quad a=1\ldots n\,,\nonumber
\end{align}
the last of the above forms showing that it is  an ensemble average of $d^{(a)}$.

By linearity, the ensemble densities satisfy 
\beq\label{ens_conservation}
\partial_t\rho_\text{ens}=-\sum_{i,a}\partial_{i_a}j_{\text{ens}\,i_a}\,,
\enq
\begin{align}\label{ensNewton}
\frac{\d}{\d t} \Big\langle p_{\text{ens}\,i_a}\Big\rangle_{{\mathbb R}^{3n}} =& -\Big\langle\partial_{i_a} V\,\rho_\text{ens}\Big\rangle_{{\mathbb R}^{3n}} +q^{(a)}\Big\langle E_i\left({\boldsymbol x}^{(a)},t\right)\, \rho_\text{ens}\Big\rangle_{{\mathbb R}^{3n}}\\& + q^{(a)}\Big\langle \epsilon_{ikl}\,\lsuper{e}j_{\text{ens}\,k_a}B_l\left( {\boldsymbol x}^{(a)},t\right) \Big\rangle_{{\mathbb R}^{3n}} +f_{\text{ens}\,i}^{(a)}(t)\nonumber\,,
\end{align}
\beq\label{ens_energy_density}
\frac{\d}{\d t}\Big\langle\,\varepsilon_{\text{ens}\,a}\Big\rangle_{{\mathbb R}^{3n}}=q^{(a)}\Big\langle E_i\left({\boldsymbol x}^{(a)},t\right) j_{\text{ens}\,i_a}\Big\rangle_{{\mathbb R}^{3n}}
-q^{(a)}\Big\langle \partial_{i_a}V\, j_{\text{ens}\,i_a}\Big\rangle_{{\mathbb R}^{3n}}
+f_\text{ens\,0}^{(a)}\,.
\enq
Integrating \eqref{interaction_energy} over $e$ with respect to $\d\mu(e)$ we further get
\beq\label{ens_interaction_energy}
\Big\langle\Theta_\text{int}^\text{ens}{}^{00}\Big\rangle=\Big\langle V\,\rho_\text{ens}\Big\rangle_{{\mathbb R}^{3n}}+\sum_a q^{(a)}\Big\langle j_{\text{ens}\,i_a}A_i\left( {\boldsymbol x}^{(a)},t\right)\Big\rangle_{{\mathbb R}^{3n}}+ q^{(a)}\Big\langle \rho_\text{ens}\,\varphi\left( {\boldsymbol x}^{(a)},t\right)\Big\rangle_{{\mathbb R}^{3n}}\,.
\enq 
As in the single particle case the functions $f_{\text{ens}\,\mu}^{(a)}=\int\d\mu(e)\, \lsuper{e}f_\mu^{(a)}$, are assumed to vanish due to  the incoherent superposition of the different $\lsuper{e}f_\mu^{(a)}$ in it.

The ensemble densities $p_\text{ens}$ and $\varepsilon_\text{ens}$, are only defined by our system up to a $3n$ divergence vanishing at spatial infinity plus a time-independent constant (-vector in the $p$ case). To fix their point wise value, we resort to two Newtonian provisos: 
\beq\label{Newtonian_proviso}
( \text{a})\quad p_\text{mar}^{(a)}=m^{(a)}j_\text{mar}^{(a)}\,,\qquad (\text{b})\quad \varepsilon_\text{mar}^{(a)}\geq 0\,,
\enq
where $m^{(a)}$ is the mass of a particle, a concept which is missing altogether from the basic tenets. 
Proviso (a) states that, at least on average, each  particle has a fixed mass-to-charge density ratio, whereas (b) is the usual positivity requirement from the $\epsilon^{(a)}$'s, and the arbitrary constant is taken as the rest energy of the particle.

\subsubsection{The many-body Schr{\"o}dinger equation}
A systematic way of generating densities $\rho_\text{ens}$,  $j_\text{ens}$,  $p_\text{ens}$  and $\varepsilon_\text{ens}$, satisfying the (highly non trivial!) $f_\text{ens}$-free system \eqref{ens_conservation}--\eqref{ens_energy_density}+\eqref{Newtonian_proviso}, and consistently transforming under rotations, is via normalized  spinor solutions of the multi-particle Schr{\"o}dinger's equation. The simplest case is the scalar (spin-$0$) equation but for further use we shall use the spin-$\half$  equation  $i\hbar\partial_t\phi={\cal H}\phi$, with Hamiltonian ${\cal H}=V+\sum_a {\cal H}^{(a)}$, where
\beq\label{spinH}
{\cal H}^{(a)}= \frac{1}{2m^{(a)}}\Big( -i\hbar{\boldsymbol \nabla}^{(a)}-q^{(a)}{\boldsymbol A}\left({\boldsymbol x}^{(a)},t\right)\Big)^2+\text{g}^{(a)}{\boldsymbol \sigma}^{(a)}\cdot{\boldsymbol B}\left( x^{(a)},t\right)+\varphi\left({\boldsymbol x}^{(a)},t\right)\,,
\enq  
with $\sigma_i$, $i=1\ldots3$,  Pauli matrices,  ${\text g}^{(a)}$ arbitrary constants and $V$ the inter-particle Coulomb potential \eqref{mppotential}.  More specifically,  $\phi$ is a $H_\text{space}\otimes H_\text{spin}$-valued function of time, where $H_\text{space}$ is the Hilbert space of square-integrable functions of $3n$ variables and $H_\text{spin}=H^{(1)}\otimes\cdots\otimes H^{(n)}$ with $H^{(a)}={\mathbb C}^2$ the spin space of particle $a$.  Each spin term appearing in \eqref{spinH} acts only on the relevant spinor space and should more accurately read: ${\mathbf 1}_\text{space}\otimes{\mathbf 1}^{(1)}\otimes\cdots\otimes{\boldsymbol \sigma}\cdot{\boldsymbol B}\left( x^{(a)}\right)\otimes\cdots\otimes{\mathbf 1}
^{(n)}$. 

As is customary, it is more convenient to view $\phi$ as a $H_\text{spin}$-valued function of ${\mathbb R}^{3n}\times {\mathbb R}$ ($3n$ configuration space and time). We can then define $\phi^\dagger$ as its dual in $H_\text{spin}^*$, and with these definitions  our desired, real valued  ensemble densities are given by 
\begin{align}\label{Schrodinger_densities}
& \varepsilon_{\text{ens}\,a}=\frac{1}{2m^{(a)}}\Big[\Big( -i\hbar\partial_{i_a}-q^{(a)} A_i\left({\boldsymbol x}^{(a)},t\right)\Big)\phi\Big]^\dagger\Big( -i\hbar\partial_{i_a}-q^{(a)} A_i\left({\boldsymbol x}^{(a)},t\right)\Big)\phi\geq0\,,\nonumber\\
& j_{\text{ens}\,i_a}=\frac{1}{m^{(a)}}\text{Im}\,\phi^\dagger\left(\hbar\partial_{i_a}-iq^{(a)}A_i\left({\boldsymbol x}^{(a)},t\right)\right)\phi+\frac{\text{g}^{(a)}}{q^{(a)}}\epsilon_{ilk}\partial_{l_a}\left( \phi^\dagger\sigma_k^{(a)}\phi\right)\,, \\
&\rho_\text{ens}=\phi^\dagger\phi\geq0\,,\qquad p_{\text{ens}\,i_a}=m^{(a)}j_{\text{ens}\,i_a}\,.\nonumber
\end{align}
As in Dirac's equation,  by \eqref{ens_interaction_energy} we have
\beq\label{total_H}
\Big\langle \Theta_\text{int}^\text{ens}{}^{00}\Big\rangle+\sum_a \Big\langle\varepsilon_\text{mar}^{(a)}\Big\rangle=\Big\langle\phi^\dagger{\cal H}\phi\Big\rangle_{{\mathbb R}^{3n}} \,.
\enq
An explicit time dependence of ${\cal H}$, rendering \eqref{total_H} time dependent, is understood as a result of ignoring both  $\Theta_\text{ext}{}^{00}$ and the ensemble Poynting flux obtained by integrating the r.h.s. of \eqref{first_energy} over the ensemble.

One can see that, indeed,  only the `spin' (second) part\footnote{In the case of a single particle, the normalized magnetic moment associated with this  spin current, ${\bs \mu}=\half\int\d^3x\,{\bs x}\times {\bs j}_\text{spn}$, is easily read from $\phi$. Writing $\phi_\pm:=f(x)\chi_\pm^{({\bs n})}$ with a normalized $f$, and $\chi_\pm^{({\bs n})}$ normalized spinors defined (modulo a phase) by ${\bs n}\cdot{\bs \sigma}\chi_\pm^{({\bs n})}=\pm\chi_\pm^{({\bs n})}$; $\parallel{\bs n}\parallel=1$, one readily gets from the commutation relations of Pauli matrices that ${\bs \mu}= \pm {\bs n}$. The Bloch sphere thus obtains a very concrete physical meaning. } of $j_\text{ens}$ appears in the magnetic energy term, ${\boldsymbol j}\cdot {\boldsymbol A}$, of \eqref{ens_interaction_energy}. The self-consistency of our analysis therefore requires that the corresponding contribution of the `orbital' (first) part of $j_\text{ens}$ be much smaller. This is evidently true when our ensemble particles are entire neutral atoms, and the internal current is due to their electrons, as in the Stern-Gerlach experiment.

The omission of  `spin-spin'  interaction terms necessitates sufficiently small couplings $g^{(a)}$. This is reasonably satisfied for the Pauli choice $g^{(a)}=-q^{(a)}\hbar/\left(2m^{(a)}\right)$, representing electrons, but  a more consistent treatment, explicitly involving spin-spin interaction, requires the (multi-particle) Breit equation.

The poof that the densities \eqref{Schrodinger_densities} satisfy the $f_\text{ens}$-free system \eqref{ens_conservation}--\eqref{ens_energy_density}+\eqref{Newtonian_proviso},  basically involves a straightforward application of Heisenberg's e.o.m.'s to various operators, but \emph{crucially} depend on $(\varphi, {\boldsymbol A})$ satisfying Maxwell's equations. Moreover, Other than the assumption of slowly moving particles and the approximation  involved in ignoring the spin-spin interaction, our `derivation' of  the many body Schr{\"o}dinger equation involved no further approximations. In particular, the inter-particle Coulomb interaction is treated exactly, covering (realistic) scenarios in which the particles are not spherically symmetric, very close to each other compared with their size, and possibly even overlapping. 

Finally, it is well known that inter point-particle potentials other than the Coulomb one regularly appear in Schr{\"o}dinger's equation. Insofar as their use is warranted, they should be viewed as emergent effective potentials rather than another form of interaction. For example, the d.o.f.'s of the frenetic electrons in a lattice, holding the slow and heavy nucleons in place, can be integrated out, leaving the effective quadratic potential between neighboring  nuclei   which is used to describe lattice vibrations.

\subsubsection{Example: Violations of Bell's inequality with entangled  spin-$\half$ particles }\label{sum-Bell}

To explicitly see in action the transition from a purely classical notion, viz., the basic tenets, to QM, we shall analyse bellow an experiment involving  two of the hallmarks of QM: entanglement, and non-integer spin.

Our spin correlation experiment goes as follows. We assume that each  pair of particles whose spins are to be measured, is initially in a trap, represented by an external potential, $\varphi$, and first seek a suitable wave function to represent the joint density of a single such pair while it is still in the trap. Note that the joint density in this case represents a time-average, viz., an ensemble of densities of the form \eqref{dnn} associated with a single pair, each computed from  a different time-slice. Assuming that the trapping time of each pair is sufficiently long so as to allow the particles to reach a steady state equilibrium (see section \ref{The ground state} next for a detailed explanation) we therefore seek a wave function, $\phi$, representing an equilibrium state, which is defined by the condition that its associated  joint densities \eqref{Schrodinger_densities} (hence also the corresponding marginal densities \eqref{marginal_densities}) are time-independent when averaged over a sufficiently long period. This is clearly the case for any  eigenstate of the Hamiltonian, but not only for it. A superposition of two (for simplicity)  different eigenstates results in two stationary densities---each associated with one of them---plus an oscillatory term which vanished when averaged over times much longer than the period of the oscillation. 

For a pair of particles with identical properties (mass, charge, spin...), assuming that their equilibrium state is not a static one but rather more of a  `thermodynamic' type, with both particles constantly moving and interacting (see section \ref{The ground state} below),  we should also require from the joint densities to be symmetric under swapping of particle indices $(1)\leftrightarrow (2)$ , necessitating a $\phi$ which is either symmetric or antisymmetric under such swapping.   Assuming the latter, our desired wave-function could, for example, have the form $\phi=\chi f\exp\,-iEt/\hbar$ with $\chi$ the (antisymmetric) singlet state in spinor space $H^{(1)}\otimes H^{(2)}$, $f\left({\boldsymbol x}^{(1)},{\boldsymbol x}^{(2)}\right)$ a normalized symmetric eigenstate of the spatial part of the Hamiltonian, and $E=\left\langle\phi^\dagger{\cal H}\phi\right\rangle_{{\mathbb R}^6}$ its energy. We also assume that the same eigenstate, such as the ground state, represents each and every pair, but this is not essential to the analysis; we can always treat the ensemble as a `statistical mixture' of sub-ensembles, each represented by a different eigenstate. What \emph{is} crucial is the existence of an interaction term in the the Hamiltonian, preventing from $\phi$ from factoring into a (trivial) product state $\phi_1\otimes\phi_2$.

Now that we have the form of the ensemble density associated with each pair while it is still in the trap, and that it is one and the same, we also know what the ensemble density associated with the ensemble of all \emph{different} pairs initially is: It is again, that same density. And since that density is derived from a known wave-function, we get the form of the wave-function representing our correlation experiment at times when the particles are still in the trap.  Note that the  initial two-particle wave function, not depending on the orientations of either polarimeter, reflects the initial indifference of the joint ensemble densities  to those orientations rather than the indifference of the individual currents \eqref{dnn}--\eqref{En}.  Moreover, the subsequent propagation in time of that initial wave-function,  has no physical meaning.  It is merely the way we construct the joint ensemble densities: one space-like slice at a time, further recalling that this `time index' represents different  physical times to different ensemble members, as we  had to shift them all to a common support in time, in order for our construction to work.

When the potential $\varphi$ changes and $\phi$ is allowed to propagate to regions in ${\mathbb R}^3\times{\mathbb R}^3$ containing the strong magnetic gradients inside the  polarimeters (represented by ${\boldsymbol A}$) the analysis proceeds as in QM textbooks: We  compute the joint normalized density $\rho_\text{ens}=\phi^\dagger\phi$ at regions in ${\mathbb R}^3\times{\mathbb R}^3$ corresponding to the exits from the two polarimeters. This joint density is peaked at eight different regions,\footnote{A typical peak corresponds to particle $1$ exiting polarimeter $A$ with `spin up' and particle $2$ exiting polarimeter $B$ with `spin down' which has the same integral as the peak corresponding to the above with $1\leftrightarrow 2$.} the integral over each measuring half the probability of  each of the four `spin pair measurements',   from which  spin correlation statistics can be deduced for each choice of polarimeter's  angles and---low and behold---Bell's inequality is violated! The clustering of the joint density into eight regions rather than, say, two (spin-$0$ Schr{\"o}dinger's equation)) or 18 (spin-$1$ equation) regions, is obviously a statistical property of a particular form of internal particle current---ordinary conserved electric current in our picture---and there is nothing more counter-intuitive to spin measurements than to any other QM scattering experiment.     

So why doesn't Bell's inequality apply to our ECD particles? Although working more than half a century after the  sound establishment of relativity theory, Bell's modelling of particles is entirely that of a nineteenth century physicist (not to say a Greek philosopher). A particle in Bell's analysis of our correlation experiment, would be a  small corpuscular with, at most, a tiny midget inside, guiding the particle  in strong magnetic gradients (as already remarked, in slowly varying EM fields on the scale of the particle, the Lorentz force equation follows from the basic tenets and  the midget's role becomes inconsequential). Consecutive runs of the experiment can only differ by the sort of midget sitting inside the particle, who is initially oblivious to the orientations of either polarimeter, but is free to communicate with his  midget partner while they are both  in the trap.

From a relativistic standpoint, Bell's modelling of the experiment is not even wrong. A particle in the block-universe is represented by various `world-jets', co-localized around a single world-line, each corresponding to one of the single-particle densities $\rho, j, p, \epsilon$. Repeated runs of an experiment involve an ensemble of such world-jets. And when two particles participate in each run, and those are  forced to interact (as happens in the trap in our correlation experiment example), the object which is relevant to the  analysis  becomes a `world-tree', with the trunk representing the interaction (e.g. trapping epoch),  later splitting into two branches, each ending in a different polarimeter. It is the ensemble of such trees, each giving rise to a single product joint distribution \eqref{dnn}, but collectively adding to a non product joint distribution,  which encodes the results of our correlation experiment. And what we have shown is that the many body Schr{\"o}dinger equation is a natural tool for computing statistics of ensembles of such objects, subject to the highly non trivial constraints \eqref{ens_conservation}--\eqref{ens_interaction_energy}+\eqref{Newtonian_proviso}, imposed by the basic tenets.  

One can, of course,  assign a certain space-like slice from the trunk (or the trunk as a whole)  the role of the hidden midget (variable) (although  it is doubtful that Bell would have subscribed to such an adaptations, which is rather removed from the clear Newtonian logic of his theorem), povided that such a slice encodes  sufficient information so as to determine the behaviour of each particle at its polarimeter.  In a block universe of the IVP type (see section \ref{The block universe} for a reminder) that would be the case. However, as noted before, currently the only known realization of the basic tenets \cite{YK2016} does not lead to an IVP block-universe. And even if it did, one can still seriously object to Bell's requirement that the distribution of those hidden variables should be independent of the orientations of the polarimeters. Looking at each tree as a single whole, it is only natural to expect that trees with different branch endings would also have different trunks. This is also referred to as the retro-causality case against Bell's inequality,  \cite{Wharton2} \cite{Price2}  \cite{Argaman}, which likewise relies on the block-universe view.    Finally, our analysis clearly shows that particles (or other systems for that matter) interacting in their \emph{future}, should exhibit certain correlations in the past which are unexpected from a Newtonian standpoint.  
\subsection{The Compton length}\label{Compton}
The ensemble densities associated with Dirac's equation, converge in the limit  $c\rightarrow \infty$ to those of the single particle Schr{\"o}dinger equation. Besides loosing manifest Lorentz covariance---which is completely acceptable, as our derivation of Schr{\"o}dinger's equation  was committed to one particular reference frame---a couple of other properties of Dirac's equation are lost: An intrinsic length scale,   $\hbar/(mc)$, the so-called Compton wave-length, which appears in the free Dirac equation, disappears from  Schr{\"o}dinger's, along with various bizarre phenomena associated with the the former. Next, we argue that the two effects are closely related but distinct nonetheless.

The first such bizzare phenomenon, known as Zitterbewegung, is the occurrence of ultra-high frequencies in the Dirac current \eqref{Dirac_current} associated (even) with a freely moving wave-packet which is localized on scales comparable with the Compton length $\hbar/m$ of a particle (see, e.g., chapter 2 in \cite{Itzykson}). Assuming that electron densities have a typical extent equal to the Compton length, trying to localize an ensemble of electrons on scales beneath it, clearly must lead to a paradox. The Compton length emerges therefore as the extent of the e-m density of a particle, corroborating the results of section \ref{hbar limit} on the $\hbar\rightarrow 0$ limit.  Note also that by avoiding such absurdities, and further assuming that the charge density is likewise confined to the Compton length scale, our consistency condition regarding the omission of the $f_\text{ens}$ term in \eqref{hcc} and \eqref{hcc2} is automatically satisfied (see equation \eqref{convolution}).

Artificially localizing a wave packet on a too small a scale is only one way such a contradiction can occur.  An  external field, $F_\text{ext}$, whose scale of variation is on the order of the Compton length or beneath, generally leads to a  modulation of the Dirac densities  with a comparable scale of variation. The famous Klein paradox (whose true nature is, in fact, somewhat less `paradoxical' than commonly stated; see \cite{Grubl}) is such an example where the step potential creates  an illegal modulation and, by our interpretation, has nothing to do with pairs creation at the step barrier, as commonly stated in QM textbooks.

Another example for an illegal modulation of an otherwise wide wave packet, is involved in the attempt to explain Compton's scattering based on the Dirac equation. In this case, the external potential is a plain wave with a wave length which is on the order of the electron's Compton length. This difficulty disappears for longer wave-length and the correct Thompson cross section is reproduced simply from the Poynting flux associate with $\tilde{F}_\text{ens}$.  Klein and Nishina, trying to extend this result to higher frequencies, had to abandon this purely classical analysis and resort to a deus ex machina, known as the `correspondence principle', in interpreting $\tilde{F}_\text{ens}$. 

The disappearance of the  Compton length in passing from Dirac's to Schr{\"o}dinger's equation might be interpreted as implying that the latter deals with point particles only, but this impression is wrong. On the contrary, it means that particles described by Schr{\"o}dinger's equation can have \emph{any} size. This conclusion is consistent  with our derivation of Schr{\"o}dinger's equation, making no such limitation on the size of individual ensemble members, by the point-particle-limit being---as in Dirac's case---the $\hbar\rightarrow 0$ limit (see section \ref{hbar limit}), as well as with the wide experimental scope of Schr{\"o}dinger's equation, involving anything from subatomic particles to entire molecules and BEC's (By our derivation of Schr{\"o}dinger's equation, this universal applicability supports a conjecture, to be covered in another paper, that at the fundamental level there is nothing but classical electrodynamics, viz., densities satisfying the basic tenets). 

Nonetheless,  whereas  in  Dirac's equation, incorporating an intrinsic length, we are `warned' of trying to localize the ensemble densities  on scales beneath that length scale, no such warning exists in the Schr{\"o}dinger case. And, indeed, no paradoxes of the form previously mentioned appear there.   
The caveat of not localizing a wave function on scales beneath the size of individual particles, and of not modulating it with a rapidly varying external field over that size, must therefore be insert `by hand'. Such caveats add to the more familiar one of not initializing a wave function with too narrow a support, which would lead to subsequent faster-than-light spreading.

\section{The `conspiracy' behind the invention of the `photon'}\label{conspiracy}
In the previous section we showed that single-body relativistic QM and many-body non-relativistic QM provide a plausible statistical description of a generic block-universe constrained by the basic tenets. In the current section we extend that analysis to phenomena involving EM radiation.  To maintain the level of rigour from the previous section, we would need to derive equations for ensembles of many-body relativistic particles involving also the (specific to a member of the ensemble...) EM field---a task currently underway. Nevertheless, since all we know about radiation comes from its interaction with matter, we already have at our disposal enough tools to argue the case for the `photon' being merely a (somewhat conspirational) coincidence of various statistical aspects of the ECD block-universe.

We begin by checking what the ensemble densities of section \ref{The ensemble current} can tell us about photons. A central notion in this analysis is the following: A system is said to be in \emph{equilibrium} with the global EM field if the integrated Poynting flux across a time-like three cylinder containing it, vanishes when the cylinder is taken long enough---the relevant time-scale being dictated by the system. In other words, a system is in equilibrium if, on average, it absorbs just as much energy as it radiates, maintaining this way a constant time-averaged energy. A single radiating particle in equilibrium, for example, must generate, on average, equally as much retarded flux as advanced one, or equivalently, have a nearly balanced time averaged $\langle\alpha_\text{ret}\rangle=\langle\alpha_\text{adv}\rangle=1/2$, along its current. Any system undergoing an  irreversible transition (in the thermodynamic sense), such as a burning candle, is clearly not in equilibrium.

\subsection{What is a Hydrogen atom?}
The normalized electric density, $\rho_\text{ens}(\boldsymbol{x})$, associated with an eigenstate of the Hydrogen Hamiltonian has a simple and regular form. But what is, presumably, the underlying electron normalized density, $\rho(\boldsymbol{x},t)$, related to the former via
\beq\label{ell}
\rho_\text{ens}
(\boldsymbol{x})= \lim_{T\rightarrow \infty}T^{-1}\int_0^T \d t \rho(\boldsymbol{x},t)\,.
\enq
One guiding principle in answering this question follows directly from the basic tenets (see appendix D in \cite{YK2016} for a relativistic treatment): When the electron is sufficiently small on the scale of variation of an external EM field and the self-force is ignored, the (center-of-mass of the-) particle must describe classical paths of point particles in the external field. For the special case of the Coulomb potential, which is a harmonic function, this result is \emph{exact} so long as the particle maintains a spherically symmetric charge distribution. This could explain the bizarre coincidence of the classical Rutherford cross section with its QM counterpart at \emph{low energies}---the opposite of the so-called `classical limit'---and even more remarkably, with the QED Bhabah (electron-positron) and M{\o}ller (electro-electron) scattering cross sections at low energies, where the Coulomb potential does not even directly appear.  

When applied to a bound electron,  taking the relevant length scales into account (the Bohr radius and the Compton length, assumed to be the size of the electron) one might  conclude that it moves almost classically  at distances greater than the Bohr radius. However, the self-force is not negligible in this case. The EM power radiated by a point electron classically orbiting the proton at the Bohr radius is $\sim 10^{17}m_ec^2$ per orbit! In other words, at such close proximity to the nucleus,  self fields  necessarily render the electron's trajectory  highly non-Keplerian or else, even if the self-force vanished (due to the above power being equally distributed between advanced and retarded self-fields) an unrealistic  amount of EM energy would have had to surround every electron.  These self fields effects strongly depend on both the electron's (time dependent) morphology and on its associated  $\alpha$'s (equation \eqref{convolution with K}).  

%

As the electron is assumed to be in equilibrium, its energy only fluctuates around  its mean. The same applies to the magnitude of its angular momentum but as for its direction, a random walk on a sphere, of the type induced by the `self-torque', results in an isotropically pointing vector. Stationary ensemble currents having a net angular momentum, viz., those derived from $(n,l\neq0, m\neq 0)$ wave-functions, can only represent short episodes of a system  unless a magnetic field, breaking the isotropy of the system, is introduced. This means, among else, that q-bits represented by spins have a limited lifetime regardless of any specific decoherence mechanism. 

\subsection{The ground state}\label{The ground state}
\begin{figure}
	\centering
		\includegraphics[width=0.7\textwidth]{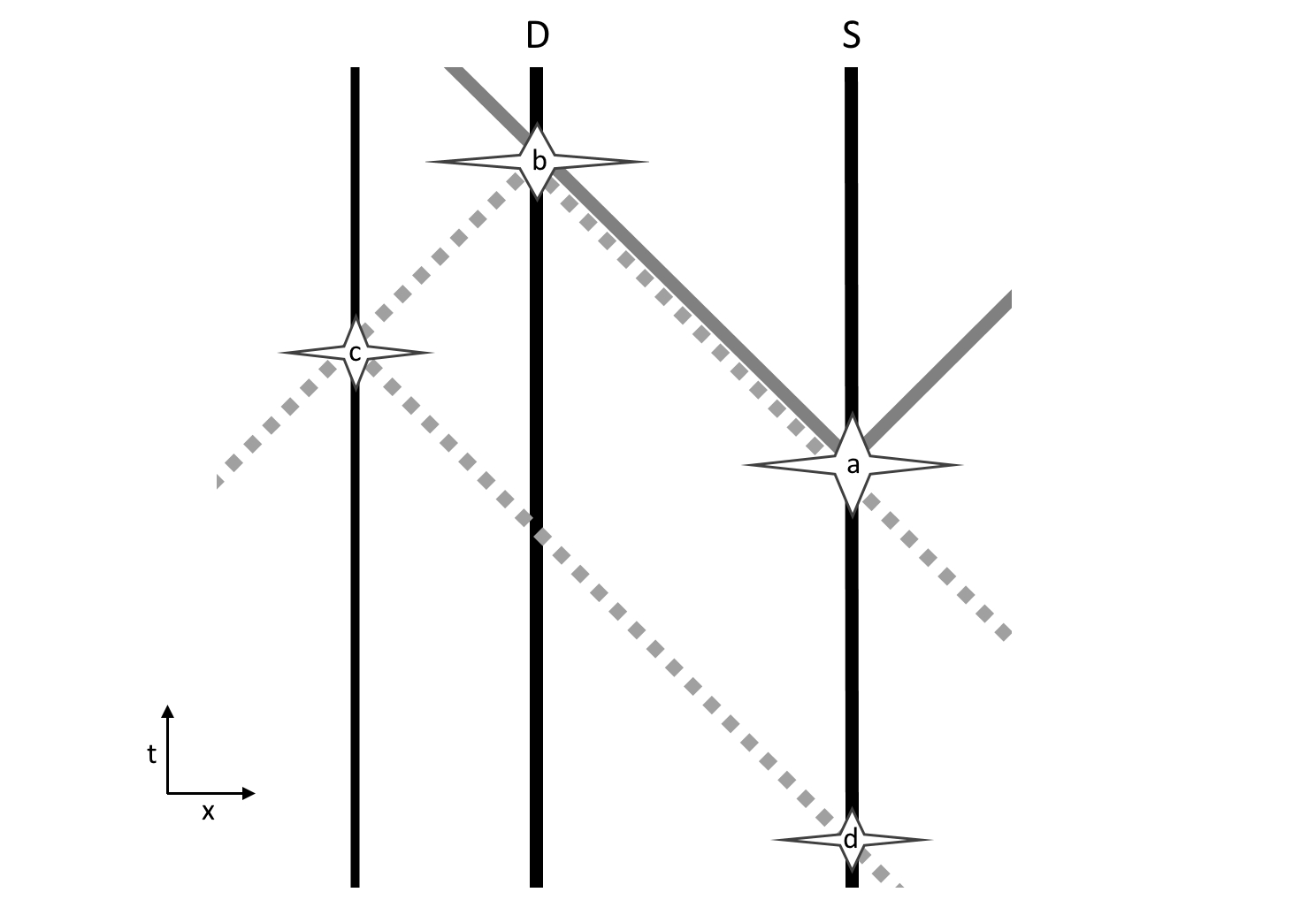}
	\caption{Particle S radiates retarded waves (solid gray lines) at point $a$ in space-time. Particle D reacts to them at a later point $b$, generating advanced waves (dashed gray line) affecting particle S directly at $a$ and also at earlier (and later; not shown) points such as $d$ via a third particle.}
	\label{fig:source-detector}
\end{figure}

Finally, we ask why the ground state is not just  stationary (extended) electrons minimizing the electrostatic energy of the system?  The answer is that, in principle, this could have been the case. However, when a system dissipates its energy into retarded waves, those waves eventually interact with other atoms in the universe---possibly in remote galaxies too. These atoms react to the retarded field---and here is the crucial point---generate also advanced fields, affecting  our radiating system directly, and also indirectly even at earlier points along its world line (see figure \ref{fig:source-detector}). 

From the above description it is  clear that  a block-universe consistent with such `feedbacks' must be in some sort of global quantum equilibrium. As noted, a particle feels an instantaneous reaction to any change in its state  when advanced fields are allowed. And although the strength of this reaction decreases as one over the distance squared between the first particle and the reacting particle, there are more such reacting particles at greater distances (in a statistically uniform universe) by that exact same factor. It is further clear that such a universe which is also consistent with the local basic tenets, must exhibit a much higher degree of coordination between remote regions than a universe containing only retarded fields, in which the reaction is delayed by twice the light-time between the particles. We shall examine below a few possible manifestations of such a coordination. 

The feable reaction field at a point $a$, coming from the rest of the currents in the universe at their intersection with the light-cone of $a$, which are responsible for the nonvanishing mechanical energy of the ground state, will be refered to as the \emph{zero point field} (ZPF), a name we borrow from stochastic electrodynamics\footnote{A somewhat  related picture of an atom is given by  stochastic electrodynamics (SED; see e.g. \cite{SED}). In  SED,  point charges generate only retarded waves, self-force effects are (allegedly) absorbed into their dynamics by replacing the Lorentz force equation with the Abraham-Lorentz-Dirac (ALD) equation, and a zero-point-field (ZPF) occasionally compensates the electron for its radiated energy loss, preventing it from spiralling into the nucleus, as otherwise follows from the ALD equation. As the name implies, the source of ZPF are retarded waves generated by the rest of the particles in the universe which are at their ground state.  

There are, however, some crucial differences between the two ZPF's. First, the SED ZPF, on average, must `pump' energy back to the electron to save it from `spiralling to death',  whereas in ECD, the electron radiates both advanced and retarded fields and needs no such salvation to begin with. The ECD ZPF is energetically neutral.

The second difference has to do with the magnitude and spectral content of both ZPF's. The ECD one is assumed feeble compared with the other fields involved in the dynamics of an atomic electron---the self-field and the Coulomb central field. It is also supposed to have an upper frequency  cutoff equal to $\hbar/(m_e c^2)$, inherited from the extended size of electrons, which are the source of the ZPF.  This should be contrasted with a point particle spiralling towards the Coulomb center, eventually attaining arbitrarily high frequencies. The compensating SED ZPF, therefore, not only is it not band limited, but it further turns out that it must diverge at high frequencies to balance the ALD dissipation. The justification commonly given to this divergence is that the spectrum of the ZPF must be Lorentz invariant. This reason seems rather contrived, given that every other aspect of our (local) universe is not Lorentz invariant.

As we shall see below, advanced fields are also absolutely mandatory in any classical account of photons. Moreover, as shown in \cite{YK2016}, classical electrodynamics based on the ALD equation is inconsistent with the basic tenets. These are not only necessary in order for a theory to be compatible with the experimental scope of CE but, as we saw  in section \ref{The ensemble current}, also to establish the compatibility of an ontology with QM. We conclude that despite being in the spirit of ECD, SED is `too simple' to offer a realistic description of micro-physics. 

Nevertheless, SED has had some impressive quantitative success in reproducing certain quantum mechanical results based on the concept of ensemble average, and it is therefore tempting to apply similar methods to ECD.  However, the ECD counterparts of those methods are not only infinitely more complicated due to the extended structure of an ECD particle, but they also expose the `deception' inherent in any alleged derivation of a statistical theory from a single system theory: One must \emph{postulate} an ensemble over which the statistics is to be computed. When the single system equations are sufficiently simple, the postulated ensemble can be compactly defined, camouflaging the fact that critical information besides the single-system equations has been added to the computation. The definition of an ensemble of ECD solutions (more accurately, of `segments' cut from the global ECD solution of the block-universe), each representing a repetition of an experiment, requires an infinity of such postulates, making manifest the status of QM as a fundamental law of nature, on equal footings with the  underlying ontology---allegedly ECD---and further explains why QM could have predated ECD (or whatever underlying theory).}. The global EM field in the vicinity of a particle can therefore be decomposed into as a  ZPF,  a self field, and possibly  a regular external field.

It is the ZPF which is presumably responsible for interference effects in scattering experiments and interferometers, mediating to a particle the surrounding space-time charge distribution $(\rho,{\boldsymbol j})$. These appear implicitly in Schr{\"o}dinger's equation through the external potentials, $(\varphi,{\boldsymbol A})$, with Maxwell's equations providing the link $(\varphi,{\boldsymbol A})\rightarrow (\rho,{\boldsymbol j})$. The consistency of this description is validated by the fact that remote potentials indeed have a negligible effect on a localized solution. Their influence on a system just adds up to the `universal part' of the ZPF, coming from the rest of the remote particles in the universe, which is responsible, among else, for the spreading  of a wave function in potential free regions.

Since the influence of the ZPF is extremely feeble compared with that of the self and external fields, special amplifying conditions are necessary to get a measurable effect. In scattering experiments it is the huge distance between the scatterer      and the detection screen; In interferometers it is the great sensitivity  of the particle's  chaotic dynamics  while in the beam-splitter---a macroscopic lattice of many scatterers with which a particle interacts numerous times before `deciding' upon an exit angle.

By our analysis, the ground state of a single Hydrogen atom in an otherwise void universe is a meaningless concept. The ground state is an attribute of the entire universe which is supposed to be in some sort of global `quantum equilibrium', inducing a common equilibrium state on identical systems---Hydrogen atoms in particular. Recalling that in ECD, being scale covariant, the scale, viz., mass of a particle is not a fixed parameter but rather a variable (like its position), it is speculated  \cite{Knoll_cosmology} that the common mass of all particles of a given specie, elementary or composite, is a result of that same global quantum equilibrium. The implications of this conjecture to cosmology \cite{Knoll_cosmology} are obviously radical, and appear to be solving the major problems in cosmology.

\subsection{Perturbed atoms. A single-body analysis}\label{pa} 
If we perturb an electron which is in equilibrium---say, by applying an external electric field, as in a gas lamp---the ensemble current needs no longer be time independent and, assuming that the perturbation is sufficiently small for simplicity,  it is derived from a normalized sum of the form
\beq\label{superposition}
\sum_\lambda e^{-\text{i}E_\lambda t/\hbar}\phi_\lambda({\boldsymbol x})\,
\enq
where $\lambda$ is a multi-quantum number and $\phi_\lambda$, $E_\lambda$ are the corresponding (bound) wave-function and energy respectively. The radiation field generated by the ensemble current can easily be shown to have only discrete frequencies of the form $\left(E_\lambda-E_{\lambda'}\right)/\hbar$. Note that the discreteness of the ensemble spectrum does not imply a similar discreteness in the spectrum of individual members of the ensemble. Those latter are still largely the unperturbed ones; The discrete peaks only represent the part of the Poynting flux generated by an electron, breaching equilibrium, and if it occurs in the ensemble current it must occur in at least a subset of the ensemble of a non-vanishing measure.

Returning to our single Hydrogen atom, assumed to be in the ground state,  let a weak EM pulse impinge on it. A standard undergrad exercise in time-dependent perturbation theory shows that after the pulse has passed, we have two possibilities, depending on the nature of the pulse: If the pulse is (wide-band) random, the ensemble current hardly modifies.  	But if the pulse is centered at frequency $\omega$ and has a long coherence length, then  we could be left with a modified ensemble current, no longer derived from a single eigenstate of the Hamiltonian. How should we interpret this modified ensemble current?

To answer this question, we should  distinguish between an ionization scenario, whereby the excited eigenfunctions belong to the continuous part of the spectrum, and an excitation scenario, in which they are discrete. Focusing first on the ionization scenario, where there is no overlap between the ground state and the wave-packet spanned by the excited states, the answer is evident: Some members of the ensemble continue their frenetic motion around the nucleus, associated with the ground state, while others fly away from the atom with a kinetic energy which can be shown to be centered around $\hbar\omega+E_0$. Clearly, no electron is in a `superposition' of ionized and ground states, and  when the sub-ensemble of  ionized electrons is subsequently used in, say, a scattering experiment, it constitutes a fraction of the original ensemble consistent with the standard collapse postulate of QM. 


At this point one might conclude that the ensemble current in the above case tells us nonsense. Indeed, if the EM pulse could be arbitrarily weak and still lead to ionization, then where did the energy needed to jolt an electron come from?  It is precisely this energy conservation concern which pushed Einstein to his radical explanation of the photoelectric effect, namely, that the ionization is due to `light corpusculars' of energy $\hbar\omega$, kicking electrons out of a well of energy $E_0$. 

Nevertheless, when advanced fields are part of the analysis, as is the case in ECD, the above argument against the ensemble current is invalid. The energy needed to kick an electron out of an atom could come from advanced fields, converging on the electron (point $b$ on the world line of particle D; see figure \ref{fig:source-detector}) and yet, consistently leave no trace in the corresponding relativistic single-body QM calculation (recall how self-fields, including advanced ones, are removed from the equations for the ensemble current in section \ref{The ensemble current}).

Increasing the intensity of the EM pulse entering the calculations of the ensemble current, by our interpretation, only shifts us to a different ensemble of ECD solutions in which ionization is more frequent. Note that an additional result of that undergrad exercise is that the fraction of jolted electrons is proportional to the amplitude squared of the pulse. This means that the probability of ionizing the atom drops as one over the distance squared between the source of the pulse and the atom (as if a flux of particles is emerging from the source), again, consistent with experiments.
 
Returning to the excitation scenario, the situations is  similar.  As in the ionization case, an excitation of a bound eigenstate of the Hamiltonian, of energy $E_\text{e}$, can occurs for $\hbar \omega= E_\text{e}-E_0$. However, due to the overlap between the supports of the  bound and excited states in \eqref{superposition}, the underlying ensemble cannot consistently be assumed to be comprised of members in  \emph{either} ground or exited stationary states. Instead, it consists of members all of which are in a common, `hybrid' equilibrium. When subsequently applying an ionization EM pulse, the story repeats as in our previous, ionization case, only this time with three, non-overlapping wave-functions: Two belonging to the continuous part of the spectrum, and one being the discrete, hybrid bound state.   



\subsection{Single photon sources}
The classical account of photons ignored, thus far, the source of the EM pulse entering the Hamiltonian, involved in the computation of the ensemble current. When this source is a decaying atom, or some other simple few-body system referred to as a `single photon source' \cite{Photon}, the illusion of a real light particle becomes yet  more convincing. In this case, the single-body/system analysis is not entirely adequate, as global e-m conservation must also be taken into account. 

Figure \ref{fig:source-detector} depicts the e-m distribution allegedly associated with a single photon production-detection sequence. Atom $S$(-ource), initially in its ground state, is struck by---say---a femtoseconds laser pulse (not shown in picture).  Following a transient period, the atom finds itself in a modified equilibrium, whose ensemble current is derived from an excited eigenfunction of the atom's Hamiltonian. Why, then, does the electron suddenly fall back to a lower energy equilibrium state? The reason is clearly not the ZPF which is responsible for the atom being in equilibrium from the outset. As we saw, it takes a very specific narrow band pulse for that.

The source of such a special pulse is easily read from figure \ref{fig:source-detector}. It is the advanced field generated by the $D$(-etection) particle, triggering the decay of $S$. But the opposite direction is just as true: It is $S$'s retarded field triggering the excitation of $D$. In the block-universe, there is no cause nor effect; A process is just there, `written' in the corresponding e-m structure (distribution).

To check that the above description is consistent with our previous single body account of photons, we note from that analysis that for $D$ to gain an $\hbar\omega$ `quantum of energy', the retarded pulse generated by $S$ must be centred at $\omega$. But for this to hold true, $S$ must drop to a new equilibrium state which is $\hbar \omega$ lower. This, in turn will happen if the advanced wave generated by $D$ is, likewise, centred at $\omega$ (note that the external pulse entering the single body analysis is not `marked' by either advanced or retarded labels), which---finally closing the consistency loop---is a consequence of $D$ adding $\hbar \omega$ to its energy. Such  emissions accompanying the decay of  atomic systems are commonly referred to as `spontaneous', to distinguish them from emissions `stimulated' by a retarded pulse. But from the block-universe perspective, there is nothing more spontaneous in spontaneous emission than in stimulated emission. 
\subsection{Energy-momentum conservation joins the conspiracy} 
A system which is in equilibrium with the ZPF maintains a constant e-m only when averaged over a sufficiently long time. And since e-m is identically, locally conserved, the fluctuations in the system's e-m must be compensated by opposite fluctuations in the e-m content of the ZPF. 
When an atom which is initially in equilibrium with the ZPF, drops to a new equilibrium state whose (time-averaged) energy is lower by $\Delta E=\hbar\omega$, re-equilibrating with the ZPF, and before this energy appears in an absorbing atom, a (time averaged)  $\hbar\omega$ increase  appears in the energy content of the EM field. Now, we already know that in the process of dropping to a lower energy level, an atom releases a retarded EM pulse centred at $\omega$, and so, by the linearity of Maxwell's equations, we get that the $\hbar\omega$ energy increase is composed of a stable part, co-localized with the pulse, and a fluctuating part, coming from the ZPF.      It follows that when the pulse  is later completely  absorbed\footnote{The inclusion of advanced fields requires a modification to the notion of `complete absorption' which otherwise means that no retarded flux appears on the surface of a sufficiently large sphere containing the relevant system. The natural generalization is to postulate that no time-averaged imbalance between advanced and retarded fluxes appears across the sphere, thereby guaranteeing the conservation of e-m inside the sphere.} by some media---a bubble chamber for concreteness---following a re-equilibration period, an $\hbar\omega$ energy increment appears in it, and we ask: what form could this quantum increment take? One obvious option is that the medium heats up, roughly in correlation with the intensity of the (retarded) incident Poynting flux. However, recalling from our previous discussion that for the  source particle to decay, it needs to be `triggered' by the advanced wave of a detection atom which gains an energy quantum of $\hbar \omega$,  such distributed thermal excitation cannot be the only mode of absorption. Assuming for simplicity that the ionization energy of atoms comprising the medium is negligible compared with $\hbar \omega$, and that any jolted electron remains in free motion long after the medium re-equilibrates, we conclude that, at most, a single ionization event has occurred, in which case no significant heat is generated anywhere else in the medium, or else energy conservation is violated. This could easily be interpreted by our forensic observer, later seeing only a single track in the chamber, as `single photon detection'. 

At this point one may rightfully wonder how is it that the EM energy of the retarded pulse, widely distributed prior to the `photon detection', suddenly localises itself? The answer is that this localization process is not instantaneous but, instead, the result of the absorbing medium re-equilibrating after gaining a quantum of energy. It is essentially the same feedback, consisting of advanced and retarded waves, previously discussed in the context of the global quantum equilibrium.

In a Hanbury-Twiss type `single photon interference' experiment, e.g.  \cite{Grangier}, the retarded field generated in the initial decay could first be scattered by additional particles, comprising the mirror and beam-splitters,  creating this way an interference pattern in its intensity, and still lead to a single detection event. By our previous  single-body analysis, the number of detections in repeated runs of the experiment would match that irregular intensity pattern, leaving the experimenter rightfully wondering: What am I  seeing here, particles---or waves, after all?  

Extending the above analysis to momentum conservation as well, one can expect to obtain a similar explanation of Compton's effect. Regrettably, we already saw in section \ref{The ensemble current} that relativistic wave-equations containing an external potential whose scale of variation is on the order of the electron's (Compton) length scale, cannot consistently be used to construct ensemble currents (and indeed, the result they give in this domain defies a meaningful interpretation). Nonetheless, we can still imagine advanced waves converging on an electron, jolting it, and in the process also generating retarded waves, with a different spectral content and intensity so as to locally conserve e-m. This secondary retarded wave then triggers the jolting of yet another electron, and so on. Eventually, when all e-m of the incident EM pulse has been converted into mechanical energy of jolted electrons, their collective e-m equals to those of the pulse. 

Finally, we mention that our analysis focused on, possibly, the simplest phenomena involving photo-generation and detection. A slightly more involved case, for example, is when the decaying system generates a pulse of frequency $\omega$, but energy content other than $\hbar\omega$, say $2\hbar\omega$. In this case, two systems of excitation energy $\hbar\omega$ could be excited by a single pulse.
However, the re-equilibration of the absorbing medium now
entails the redistribution of the pulse’s e-m between two, possibly remote sites. Each
run of the experiment therefore involves a different space-time e-m distrubution---the counterpart of the space-time tree from our previous encounter with Bell in section \ref{sum-Bell}---and it just happens that Bell's inequality is violated for the ensemble thereof. In a block-universe containing also advanced radiation, adherence to Bell's inequality would be just as (un-)mysterious.  \footnote{Use of advanced fields in order to explain the non classical statistics exhibited by photons, later receiving the name `the transactional interpretation of QM", was made by Cramer in \cite{Cramer}. The construction of the global e-m distributions in that proposal uses time symmetric action-at-a-distance electrodynamics  \cite{Wheeler1}, but with self interaction naturally included (this makes mathematical sense for extended particles only).  While not being explicit about the precise physical meaning of ``hand shakes" between the detecting particles and the source, it definitely provides a conceptual way of saving locality. However, having a fixed, $\alpha_\text{ret}=\alpha_\text{adv}=\half$ division between advanced and retarded fields, his proposal cannot possibly describe photodetection. The EM field close to a charge is entirely dominated by the self field, hence without a temporary imbalance  between advanced and retarded parts, of the type appearing in ECD, no sudden delivery of energy to a charge can occur, which is mandatory in photodetection.}

\subsection{Black-body radiation}
Planck's derivation of the black-body's spectrum, seemingly requiring the `quantization of the EM field', can  be given some justifications within our current analysis, but one must admit that it could apply---at most---to cavities with nearly ideal reflecting walls. What about hot potatoes, flames, and all those other  objects exhibiting a black-body spectrum?

Sixteen years following the publication of Planck's paper which heralded the quantum era, an alternative derivation \cite{Einstein-BB} of the spectrum was proposed by Einstein, which fits like a glove to our analysis. The (classical) EM field, according to Einstein, only serves to exchange energy between particles comprising the hot body. Those, in turn, must have the standard energy distribution of a system in equilibrium. A simple compatibility condition then arises between that distribution and the spectral distribution of the EM field, leading to Planck's law. According to Einstein, atoms undergo two kind of transitions, both discussed in the current paper: so-called `spontaneous', which in our picture is just an exchange an $\hbar\omega$ `energy quanta' between atoms, and induced, either emission or absorption, where the global EM field due to all atoms plays the role of the external perturbing field (section \ref{pa}).

Einstein makes no distinction between retarded and advanced fields, so care must be practised when translating his derivation to the language of the current paper. In particular, since his derived spectral density ($\rho$) is supposed to represent the \emph{measured} spectrum, by our arrow-of-time interpretation this spectrum only samples the retarded flux, responsible for breaching quantum equilibrium. Once this is clear, the inclusion of a spontaneous emission term ($A_m^n$) but not an equal `spontaneous absorption term', is justified by the fact that the former is triggered by advanced fields, requiring special treatment, while the latter is triggered by retarded fields, hence its inclusion would `double count' the transition, which is already included in the induced transitions.

\section{Conclusion}
It is often said that the first quarter of the twentieth witnessed  two revolutions in physics---relativity theory and QM---while the remaining three quarters saw a (futile) struggle to settle their mutual conflicts. The current paper tells a different story: Both relativity and QM are partial completions of a previous incomplete revolution---Maxwell-Lorentz classical electrodynamic (CE). Einstein's relativity was born out of a need to render CE compatible with the Galilean principle of relativity; QM, we argued, is an essential complementary statistical theory of a well defined CE, free of a self-force problem. As such, not only are the two compatible but, moreover, as our analysis of Bell's inequality in section \ref{sum-Bell} shows, QM doesn't even make sense without relativity.

Obviously, there must be a third, independent completion of CE, namely, well defined CE itself, dealing with individual solutions rather than with their statistics. ECD \cite{YK2016}, we conjecture, is that theory, and there are strong indications, to be covered in another paper, that it is sufficient to describe elementary particles. That is, nothing beyond CE is needed in order to describe the subatomic world. The universal applicability of Schr{\"o}dinger's equation, from subatomic particles to complex molecules,  we briefly argued in  section \ref{Compton}, supports this conjecture, but on a more fundamental level---why should we stop trusting CE just because we move to scales much smaller than our own? Is it not a blatant repetition of the same anthropocentric bias which, time and again, has led to erroneous conclusions?  ECD, being scale covariant, is applicable to an arbitrary scale.    

Finally, we mension that although our analysis was carried entirely in flat space-time, ECD admits a straightforward curved space generalization (see section 3.4 in \cite{YK2016}). In fact, by simply adding to the basic tenets the requirement of general covariance, one is basically led to Einstein's field equations with the e-m tensor $P$ appearing in \eqref{pp} as source. Not only does this solve the self-force problem of GR, but it also has the benefit of  offering  possible conceptual foundations to quantum gravity: Just like flat space QM, quantum gravity is no more than a statistical description of the GR block-universe, encoding the statistical properties of ensembles of `segments' cut from it, each having a well defined, objectively existing  metric (not that author understands what it means for space-time not to have an objectively existing metric, a statement often made in relation with quantum gravity).\\
\linebreak
\noindent {\bf Acknowledgements}\\ The author wishes to thank an anonymous referee  for his critical remarks on an earlier draft of this paper, which greatly improved  the final outcome.

\end{document}